\documentclass[structabstract]{aa}
\usepackage[dvips]{graphicx}
\usepackage{txfonts}
\usepackage{color}
\usepackage{natbib}
\bibpunct{(}{)}{;}{a}{}{,} 
\usepackage{psfrag}
\usepackage{version}

\begin{document}

\title{Mantle formation, coagulation and the origin of cloud/core shine: \\ I. Modelling dust scattering and absorption in the infra-red}
\titlerunning{Mantle formation, coagulation and cloud/core shine}  
 
    \author{
    A.P. Jones\inst{1} \and 
    M. K\"ohler\inst{1,2} \and 
    N. Ysard\inst{1} \and 
    E. Dartois\inst{1} \and 
    M. Godard\inst{3} \and 
    L. Gavilan\inst{1}}
           
    \institute{
    Institut d'Astrophysique Spatiale, UMR8617, CNRS/Universit\'e Paris Sud, Universit\'e Paris-Saclay, Universit\'e Paris Sud, Orsay F-91405, France\and
    School of Physics and Astronomy, Queen Mary, University of London, 327 Mile End Road, London, E1 4NS, UK\and 
    Centre de Sciences Nucl\'eaires et de Sciences de la Mati\`ere, UMR8609, CNRS/Universit\'e Paris Sud, Universit\'e Paris-Saclay, Universit\'e Paris Sud, Orsay F-91405, France\\[0.1cm]
    \email{Anthony.Jones@ias.u-psud.fr} }

    \date{Received  2015 / Accepted  2016}

   \abstract
{The observed cloudshine and coreshine (C-shine) have been explained in terms of grain growth leading to enhanced scattering from clouds in the J, H and K photometric bands and the Spitzer IRAC 3.6 and 4.5\,$\mu$m bands.}
{Using our global dust modelling approach THEMIS (The Heterogeneous dust Evolution Model at the IaS) we explore the effects of dust evolution in dense clouds, through aliphatic-rich carbonaceous mantle formation and grain-grain coagulation.}
{We model the effects of wide band gap a-C:H mantle formation and the low-level aggregation of diffuse interstellar medium dust in the moderately-extinguished outer regions of molecular clouds.}
{The formation of wide band gap a-C:H mantles on amorphous silicate and amorphous carbon (a-C) grains leads to a decrease in their absorption cross-sections but no change in their scattering cross-sections at near-IR wavelengths, resulting in higher albedos.}
{The evolution of dust, with increasing density and extinction in the diffuse to dense molecular cloud transition, through mantle formation and grain aggregation, appears to be a likely explanation for the observed C-shine.}


\maketitle

\section{Introduction}

The study of starlight scattering by dust dates back to the pioneering work of \cite{1936ApJ....83..162S}, \cite{1937ApJ....85..194S}, \cite{1941ApJ....93...70H}, \cite{1968ApJ...152...59W} and \cite{1970A&A.....8..273M,1970A&A.....9...53M} who made some of the earliest measurements of the rather high dust albedo ($a \gtrsim 0.6$) and its scattering asymmetry parameter at visible wavelengths. Later studies showed that the near-infrared (NIR) albedo in the J ($1.2\,\mu$m), H ($1.6\,\mu$m) and K ($2.2\,\mu$m) bands is also high ($\sim 0.6-0.8$)  \citep{1994ApJ...427..227W,1996A&A...309..570L}. Further,  \cite{1996A&A...309..570L} showed that the NIR surface brightness of translucent and denser clouds can be explained by scattered radiation with little need for a contribution from dust emission at these wavelengths. 

Motivated by the ``extraordinary'' images of the Perseus region made by \cite{2006ApJ...636L.105F},  \cite{2006ApJ...636L.101P} provided a formalism to convert NIR (J, H, and K band) scattering measurements into accurate column density estimates over a wide range of visual extinction ({\it i.e.}, $1-20$\,mag.). In their extinction mapping of the Perseus molecular cloud complex in the J, H and K bands, in regions with $A_{\rm V} < 30$\,mag, \cite{2006ApJ...636L.105F} noted some ``emission'' structures associated with the clouds, which they then dubbed ``cloudshine''. \cite{2006ApJ...636L.105F} also interpreted their observations in terms of starlight scattering by dust in clouds and assumed it to be a measure of the dust mass distribution in denser regions. 

Seemingly related to cloudshine is the observed ``emission'' in the {\it Spitzer} IRAC 3.6 and 4.5\,$\mu$m bands, which was termed ``coreshine'' by \cite{2010A&A...511A...9S}. The IRAC 5.8\,$\mu$m band, in contrast, shows weak absorption while the IRAC 8\,$\mu$m band appears in absorption. \cite{2010A&A...511A...9S} and \cite{2013A&A...559A..60A} interpret cloudshine and coreshine observations in terms of the scattering of mid-IR photons by big grains (radii $a_{\rm big} \simeq 1\,\mu$m) deeper within the cloud and conclude that this is evidence for significant dust growth within clouds.  
As \cite{2013A&A...559A..60A} find, assuming spherical and homogeneous dust particles, the required cloudshine and coreshine grains must be significantly larger than the largest diffuse interstellar medium (ISM) grains ({\it e.g.}, radii $a_{\rm ISM} \simeq 0.25\,\mu$m), from which they must have formed. This explanation requires an increase in the largest, individual grain volumes by a factor of $(a_{\rm big}/a_{\rm ISM})^3 \simeq 64$. 

The approach that we use here is an application of the composition- and size-dependent  optical properties derived for a-C(:H) materials \citep[the optEC$_{\rm (s)}$ and optEC$_{\rm (s)}$(a) datasets,][]{2012A&A...540A...1J,2012A&A...540A...2J,2012A&A...542A..98J} and the application of these within the framework of the \cite{2013A&A...558A..62J} interstellar dust model, which has been compared to recent Planck observations \citep{2015A&A...Ysard_etal,2015A&A...Fanciullo_etal}. The original \cite{2013A&A...558A..62J} dust model was revised to encompass more realistic olivine- and pyroxene-type amorphous silicates with metallic iron and iron sulphide nano-inclusions  \citep{2014A&A...565L...9K}. This model was then extended to allow us to follow the evolution of the dust properties in the transition between diffuse and dense interstellar media \citep[][and also the companion paper, {Ysard et al. 2015}, hereafter called paper II]{Faraday_Disc_paper_2014,2015A&A...0000...000}. This modelling approach provides a general framework within which the solutions to several interesting interstellar conundrums such as ``volatile'' silicon in photo-dissociation regions (PDRs), sulphur and nitrogen depletions, the origin of the blue and red photoluminescence and the diffuse interstellar bands (DIBs) may yet be found \citep{2013A&A...555A..39J,2014P&SS..100...26J}. This approach has also led to suggestions for viable routes to interstellar/circumstellar fullerene  formation \citep{2012ApJ...761...35M} and to the formation of molecular hydrogen in moderately excited photo-dissociation regions \citep{2015A&A...581A..92J}. 
Additionally, we have studied dust evolution in energetic environments such as supernova-driven shock waves and a hot coronal gas \citep{2010A&A...510A..36M,2010A&A...510A..37M,2012A&A...545A.124B,2013A&A...556A...6B,2014A&A...570A..32B}. 

In this paper we explore a likely self-consistent solution for both cloudshine and coreshine (hereafter collectively called C-shine) via dust growth through the formation of aliphatic-rich amorphous hydrocarbon mantles,\footnote{Here we imply the formation of outer a-C:H layers on grains  by the accretion of carbon and hydrogen atoms from the gas and/or by the hydrogenation of pre-existing a-C(:H) grain surfaces.} a-C:H, and the onset of grain-grain aggregation. The combination of mantle formation and minimal coagulation lead to rather unusual optical properties.  
In the companion paper \cite[][paper II]{2015A&A...Ysard_etal} we use a radiative transfer model to explore the effects of this dust evolution scenario to explain the origin of C-shine.

We now refer to our evolutionary dust modelling approach, and all future developments and extensions of it, under the umbrella acronym THEMIS (The Heterogeneous dust Evolution Model at the IaS). 
For a detailed description of the THEMIS framework the reader is referred to \cite{THEMIS}.

This paper is organised as follows:  
Section~\ref{sect_acc} considers carbon mantle formation in the ISM in terms of the mantle composition and its spectroscopic and optical properties, 
Section~\ref{sect_anderson_loc} explores a-C:H mantle effects 
on the dust absorption and scattering properties, 
Section~\ref{sect_implications} discusses the astrophysical implications and  
Section~\ref{sect_conclusions} presents our conclusions. 

\section{a-C:H formation}
\label{sect_acc}

In the ISM the amorphous hydrocarbon, a-C(:H), dust component is most likely a mix of wide band gap, H-rich, aliphatic-rich (hereafter a-C:H) and narrow band gap, H-poor, aromatic-rich (hereafter a-C) materials \citep{2012A&A...540A...1J,2012A&A...540A...2J,2012A&A...542A..98J,2013A&A...555A..39J}.\footnote{Using Figs. 1 and 5 from \cite{2012A&A...540A...2J} we can better define the a-C(:H) material descriptors ``H-rich and aliphatic-rich'' and ``H-poor and aromatic-rich'' in terms of their hydrogen atom and sp$^3$ and sp$^2$ carbon atom fractions, $X_{\rm H}$, $X_{\rm sp^3}$ and $X_{\rm sp^2}$, respectively (where $X_{\rm sp^3} + X_{\rm sp^2} +X_{\rm H} = 1$) and the ratio $R = X_{\rm sp^3}/X_{\rm sp^2}$.  The Tauc band gap for a-C(:H) materials is given by $E_{\rm g} \simeq 4.3 X_{\rm H}$ and  these materials are clearly, \\  
H-rich and aliphatic-rich for $R \gtrsim 1.5$ and $X_{\rm H} \gtrsim 0.5$ ($\equiv E_{\rm g} > 2$\,eV) and \\ 
H-poor and aromatic-rich for $R \lesssim 0.5$ and $X_{\rm H} \lesssim 0.2$ ($\equiv E_{\rm g} < 1$\,eV). \\
Intermediate materials ($R \simeq 0.5-1.5$, $X_{\rm H} \simeq 0.2 - 0.5$, $ E_{\rm g} \simeq 1 - 2$\,eV)  exhibit $X_{\rm sp^3} \sim X_{\rm sp^2}$ and most of their sp$^2$ carbon atoms are in olefinic bonding configurations rather than in aromatic structures.} The accretion of C and H atoms in the diffuse ISM is often assumed to lead to the formation of a-C:H mantles on aromatic-rich carbonaceous dust surfaces \citep[{\it e.g.},][]{1990QJRAS..31..567J,2002ApJ...569..531M,2013ApJ...770...78C,2014ApJ...785...41C,2014ApJ...788..100C}.  The supposition of inner aromatic/outer aliphatic layering on carbonaceous dust has now been called into question in  works that point out the likely dominant effects of UV photo-processing in the diffuse ISM\footnote{Diffuse ISM is here taken to mean regions where the \cite{2013A&A...558A..62J}/\cite{2014A&A...565L...9K} model is applicable, {\it i.e.}, where A$_{\rm V} < 0.5$ and $n_{\rm H} \leq 100$\,cm$^{-3}$.}  
\citep{2011A&A...529A.146G,2012A&A...540A...1J,2012A&A...540A...2J,2012A&A...542A..98J,2013A&A...555A..39J,2013A&A...558A..62J,Faraday_Disc_paper_2014}. Further, \cite{2013A&A...558A..62J,Faraday_Disc_paper_2014} 
have shown that a-C:H mantles can probably only form in the transition from the diffuse ISM to the dense molecular clouds where there is moderate extinction \citep[$A_{\rm V} \sim 0.5 - 1.5$,][]{Faraday_Disc_paper_2014}. A self-consistent approach to the formation and (photo-)evolution of interstellar a-C(:H) materials appears to be able to explain many dust observables \citep[][paper II]{2012A&A...540A...1J,2012A&A...540A...2J,2012A&A...542A..98J,2013A&A...555A..39J,2013A&A...558A..62J,Faraday_Disc_paper_2014,2014A&A...565L...9K,2014P&SS..100...26J,2015A&A...Fanciullo_etal,2015A&A...0000...000,2015A&A...Ysard_etal}. 
 
Recent work by \cite{2012ApJ...760...36P} indicates a total interstellar carbon abundance of (C/H)$_{\rm total} \simeq 355 \pm 64$\,ppm and, with $\simeq 200$\,ppm in dust in the diffuse ISM \citep{2013A&A...558A..62J,2015A&A...Ysard_etal}, there is likely a large reservoir of gas phase carbon ($\simeq 155 \pm 64$\,ppm) in the low density ISM that could accrete onto dust in denser regions. In low extinction regions ($A_{\rm V} \lesssim 0.5$) a-C(:H) grain mantles will be composed of a-C because of photo-processing/dehydrogenation by EUV-UV photons  \citep{2012A&A...540A...1J,2012A&A...540A...2J,2013A&A...558A..62J,Faraday_Disc_paper_2014}. However, with higher density and extinction the surface layers should be be richer in hydrogen ({\it i.e.}, a-C:H) because of the reduced effects of EUV-UV photolysis and/or the increased importance of carbon mantle re-hydrogenation  \citep{2002ApJ...569..531M,2011A&A...529A.146G,Faraday_Disc_paper_2014}. Thus, C and H atom accretion from the gas phase and/or surface hydrogenation in a moderately extinguished medium ($0.5 \lesssim A_{\rm V} \lesssim 2$) could lead to the formation of a-C:H mantles on all grains. The characteristic extinction properties of this material are a strong FUV extinction and a weak UV bump, as observed along the line of sight towards \object{HD\,207198}, which shows a high carbon depletion \citep[{\it i.e.}, (C/H)$_{\rm dust} = 395 \pm 61$\,ppm,][]{2012ApJ...760...36P}. 

\begin{figure}
 \resizebox{8.0cm}{!}{\includegraphics{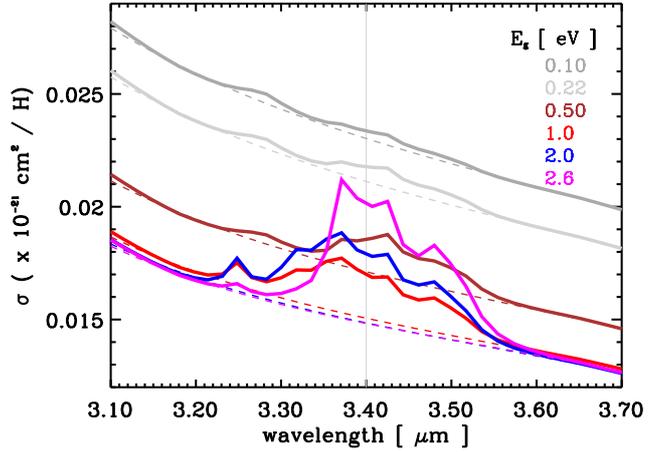}}
 \caption{The \cite{2013A&A...558A..62J} dust model extinction cross-section as a function of the outer a-C(:H) mantle band gap. The dashed lines show the adopted 4$^{\rm th}$ order polynomial baselines fitted to the model data at $\lambda =$ 3.20 and 3.65\,$\mu$m.}
 \label{fig_3mic_ext}
\end{figure}

\begin{figure}
 \resizebox{8.0cm}{!}{\includegraphics{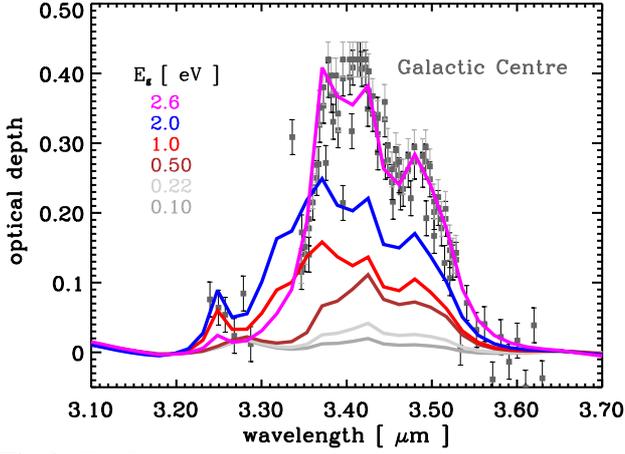}}
 \caption{The $3-4\,\mu$m region continuum-subtracted optical depth data from Fig.~\ref{fig_3mic_ext}.  Also shown are the scaled-to-fit spectra of the Galactic Centre towards \object{IRS6E} and \object{Cyg OB2 No. 12} \citep[grey squares,][]{2002ApJS..138...75P}.}
 \label{fig_3mic_spectrum}
\end{figure}

\subsection{The spectroscopy of a-C(:H) materials in the 3\,$\mu$m region } 
\label{sect_3mic_spectra}

The absorption spectra of  a-C(:H) materials in the $3-4$\,$\mu$m wavelength region show a broad, multi-component $3.2-3.6\,\mu$m feature, which is composed of about ten sub-features due to aliphatic, olefinic and aromatic C-H stretching bands. More than half of these bands are due to aliphatic CH$_n$ ($n = 1,2,3$) bonds. The detailed shape of this absorption band is therefore sensitive to the a-C(:H) material composition (H atom content and band gap, $E_{\rm g}$) and structure (C atom sp$^3$/sp$^2$ bonding ratio) and likely varies with the degree of photo-processing and/or hydrogenation experienced in the ISM \citep{2012A&A...540A...1J,2012A&A...540A...2J,2012A&A...542A..98J,2013A&A...558A..62J}. 
In Fig.~\ref{fig_3mic_ext} we show the \cite{2013A&A...558A..62J} dust model spectrum in the $3.1-3.7$\,$\mu$m region as a function of the a-C(:H) mantle composition. As per \cite{2012A&A...540A...2J} we use the Tauc definition of the band gap, $E_{\rm g}$, to characterise the a-C(:H) materials considered here. 

In addition to EUV photons, energetic heavy ion impacts, analogues of cosmic ray irradiation, can also induce structural changes in a-C(:H) solids. Ion irradiation does lead to a significant decrease in the $3-4$\,$\mu$m band strength but its effects on the constituent band-to-band ratios are seemingly small \citep{2011A&A...529A.146G}. From their study \cite{2011A&A...529A.146G} concluded that cosmic ray processing is subordinate to EUV photolytic processing in the ambient ISM except in dense clouds where the EUV photon flux is highly extinguished. 

In Fig.~\ref{fig_3mic_spectrum} we show the $3-4\,\mu$m region spectrum for the \cite{2013A&A...558A..62J} model compared to that of the diffuse ISM towards the Galactic Centre \citep[grey,][]{2002ApJS..138...75P}.

\subsection{The mid-IR properties of a-C:H}
\label{sect_MIR}

\begin{figure}
 \resizebox{\hsize}{!}{\includegraphics{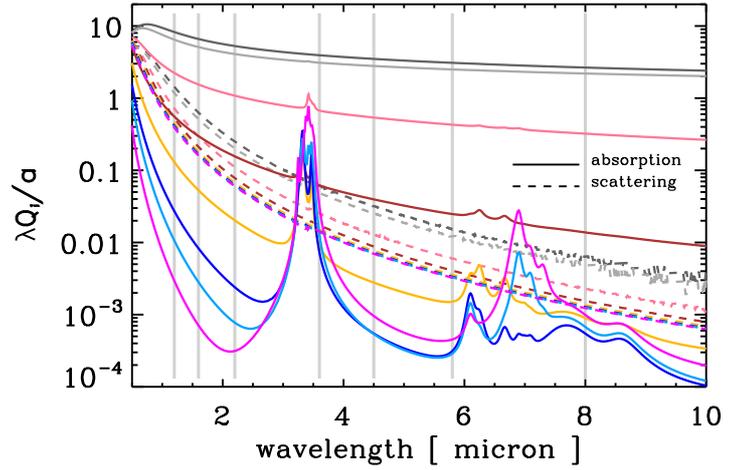}}
 \caption{The optEC$_{\rm (s)}(a)$ data plotted as $\lambda Q_{i}/a$ ($a = 100$\,nm), for absorption ($i=$ abs, dashed lines) and scattering ($i=$ sca), as a function of band gap $E_{\rm g} =$ 0.0 (black), 0.1 (grey), 0.5 (pink), 1.0 (brown), 1.5 (yellow), 2.0 (blue), 2.25 (light blue) and 2.5\,eV (violet). The vertical grey lines indicate the nominal J, H, K and IRAC 3.6, 4.5, 5.8 and 8\,$\mu$m photometric band positions.}
 \label{fig_Qs}
\end{figure}

Variations in the band gap of a-C(:H) materials have interesting effects on their optical properties.  We illustrate this in Fig.~\ref{fig_Qs} where we show the absorption and scattering efficiency factors, $Q_{\rm abs}$ and $Q_{\rm sca},$\footnote{Calculated using the Mie routine BHMIE for homogeneous spheres \citep{1998asls.book.....B}.} respectively, for a-C:H particles with radius $a = 100$\,nm. For clarity we plot the data as $\lambda Q_{i}/a$ (where $i =$ abs and sca, respectively) as a function of wavelength and the Tauc band gap, $E_{\rm g} = 0 - 2.5$\,eV.  From these data we note that for 100\,nm radius a-C particles ($E_{\rm g} \lesssim 1$\,eV) the NIR extinction (1\,$\mu$m$\ \lesssim \lambda \lesssim 10 \, \mu$m) is dominated by absorption. However, for wide band gap (1\,eV $< E_{\rm g} \leqslant 2.5$\,eV) a-C:H particles the absorption decreases with increasing $E_{\rm g}$ and so allows the normally subordinate scattering to dominate at these wavelengths. 

\begin{figure}
 \resizebox{\hsize}{!}{\includegraphics{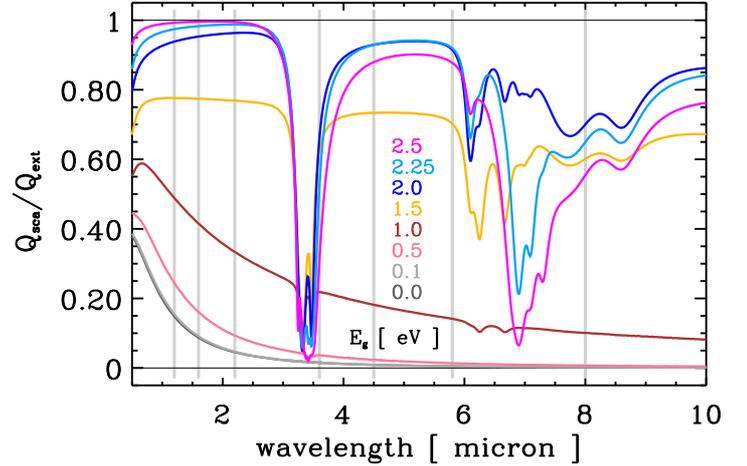}}
 \caption{The same optEC$_{\rm (s)}(a)$ data as in Fig.~\ref{fig_Qs} but plotted as $Q_{\rm sca}/Q_{\rm ext}$.}
 \label{fig_QsQe}
\end{figure}

In Fig.~\ref{fig_QsQe} we show the scattering contribution to the extinction, plotted as $Q_{\rm sca}/Q_{\rm ext}$, where $Q_{\rm ext} = Q_{\rm abs} + Q_{\rm sca}$. This figure shows that for homogeneous, a-C:H particles ($E_{\rm g} \gtrsim 2$\,eV) the J, H, K, and IRAC 4.5 and 5.8\,$\mu$m bands would be dominated by scattering. 
However, the IRAC 3.6\,$\mu$m band would be affected by the strong absorption bands in the $3.3-3.4\,\mu$m region and, more importantly, by the emission in these bands coming from stochastically-heated dust in the outer regions of the cloud. Further, the IRAC 8\,$\mu$m band would also be affected by the $7-9\,\mu$m a-C(:H) absorption bands and also by the blue wing of the $\sim 10\,\mu$m amorphous silicate band. 

The dominance of scattering in a-C:H particles is due to the fact that the absorption (scattering) is determined, principally,  by the imaginary (real) part of the complex refractive index, which becomes significantly weaker (remains almost constant) in wider band gap materials \cite[{\it e.g.},][]{2012A&A...540A...2J,2012A&A...542A..98J}. 
Thus, for a fixed mass of material (here equivalent to 100\,nm radius particles) with increasing band gap the absorption cross-section can drop by several orders of magnitude in the near-IR, whereas the scattering cross-section evolves comparatively little.

The subordination of absorption to scattering in $\simeq 100$\,nm radius, wide band gap a-C:H particles could be a manifestation of  Anderson localisation. This effect occurs in disordered (amorphous) materials due to electron wave trapping and the disappearance of conductivity in the metal-insulator transition \citep{PhysRev.109.1492},\footnote{A concept for which Philip Anderson, Nevill Mott and John van Vleck shared the 1977 Nobel Prize in Physics.} and is strongly dependent upon the dimension ({\it i.e.}, 1, 2 or 3D) and size of the disordered medium. Anderson localisation occurs with both classical light and sound waves in disordered media. Systems that scatter strongly can be synthesised but observing localisation with light is challenging because  the scattering has to be maximised without introducing absorption. This can be achieved by using light with photon energies less than the electronic band gap of a semiconductor material with a high refractive index. Additionally, when the wavelength of the photon is large compared to the particle radius ($\lambda > a$) the mean free path of the absorbing electrons, with energies near the band edge, are smaller than their equivalent wavelength and the electrons are therefore localised. Under these conditions the photons are scattered simply because they cannot be absorbed. This effect and the nearly complete localisation of near-IR light was observed by \cite{1997Natur.390..671W} in a finely-ground gallium arsenide powder. 

As well-described by \cite{1986AdPhy..35..317R}, a-C(:H) materials are disordered/amorphous solids that exhibit a wide range of semiconducting to insulating properties, depending on their band gap.  In the relatively wide band gap optEC$_{\rm (s)}$(a) a-C:H materials with $E_{\rm g} > 1.25$\,eV the refractive index, $n$, for $\lambda > 0.3\,\mu$m lies within the range $\simeq 1.69-1.76$ and is practically independent of the wavelength and the particle size \citep{2012A&A...540A...2J,2012A&A...542A..98J}, conditions that lead to almost pure scattering particles with suppressed absorption in the IR (see Fig.~\ref{fig_Qs}). These conditions are similar to those in the \cite{1997Natur.390..671W} IR scattering experiment and the observed C-shine occurring within thin layers ($\simeq 10$\,nm thick on $\sim 150$\,nm radius grains, {\it i.e.}, essentially 2D structures) of wide band gap a-C:H mantles on the grains could then be a manifestation of Anderson localisation.  

Here we considered the `ideal' and unlikely case of pure grains of wide band gap a-C:H materials. In the following section we therefore extend this study to the more realistic case of a-C(:H) layers on grain cores of various compositions. 

\section{An exploration of core/mantle grain absorption and scattering effects} 
\label{sect_anderson_loc}

We now investigate the effects of absorption and scattering within the framework of our core/mantle dust model \citep{2013A&A...558A..62J,Faraday_Disc_paper_2014,2014A&A...565L...9K,2015A&A...0000...000} and the formation of $0-30$\,nm thick carbonaceous mantles. It should be noted that the thicker mantles ($\simeq 20-30$\,nm) are for illustrative purposes only. In the ISM the carbonaceous mantles are likely to be at most $\simeq 20$\,nm thick because of cosmic carbon abundance limitations \citep{2015A&A...Ysard_etal}. To do this we explore the absorption and scattering effects of wide band gap a-C:H layers on spherical particles.

\subsection{The effects of a-C and a-C:H mantles} 
\label{sect_anderson_loc_cm}

In order to illustrate the underlying effects and to aid our analysis we begin by calculating the optical properties of core/mantle particles using the coated homogeneous sphere Mie routine BHCOAT \citep{1998asls.book.....B}. We consider grain cores of amorphous silicate grains, of olivine- and pyroxene-type composition with Fe/FeS nano-inclusions, a-Sil$_{\rm Fe,FeS}$ \citep{2014A&A...565L...9K}, and aromatic-rich amorphous carbon grains \citep{2012A&A...542A..98J} and, for each core type, we calculate the optical properties as a function of the mantle thickness, $d$, for both a-C ($E_{\rm g} = 0.1$\,eV) and a-C:H ($E_{\rm g} = 2.25$\,eV) mantles. 

\begin{figure}
 \resizebox{8.5cm}{!}{\includegraphics[angle=0]{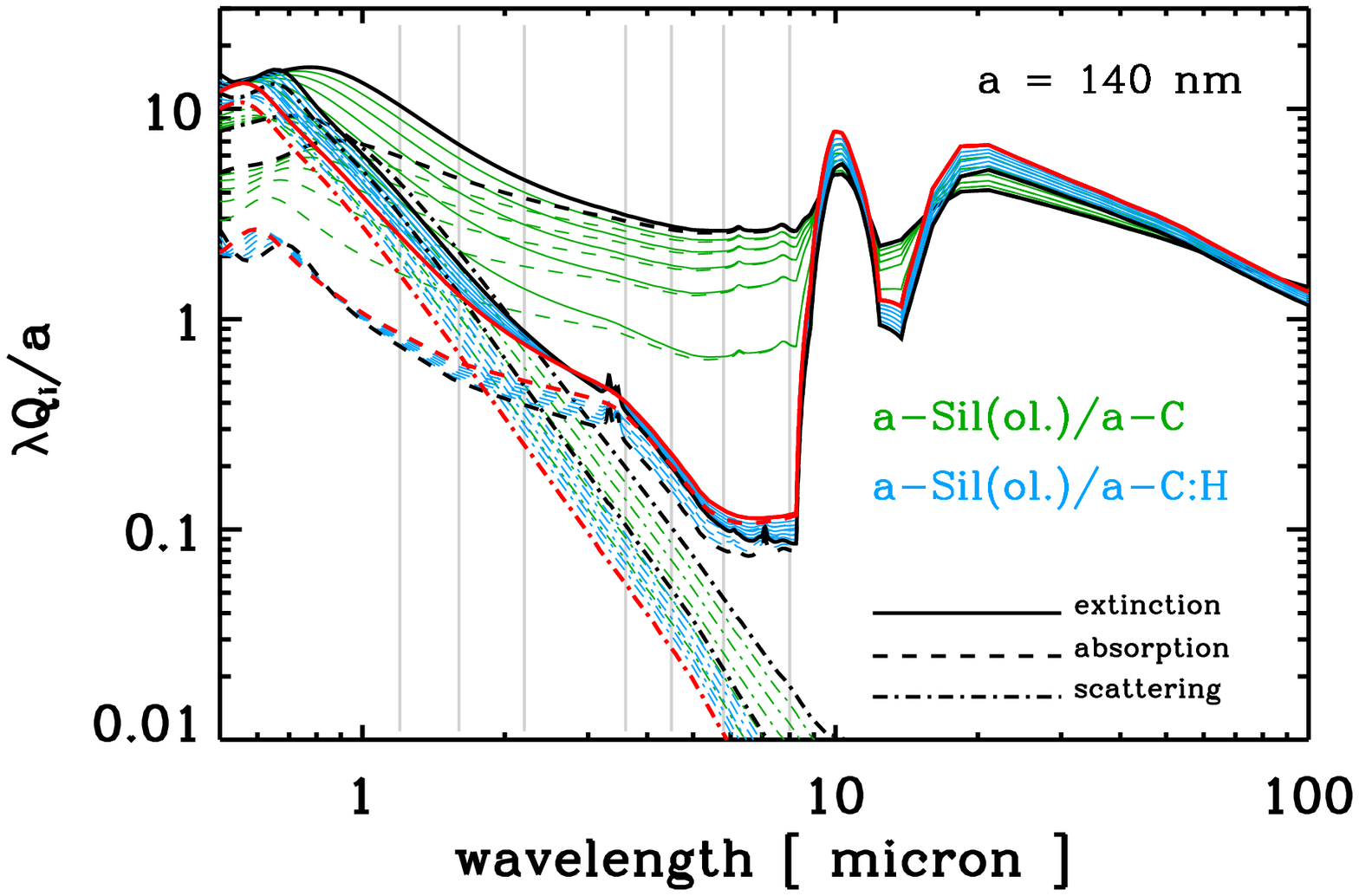}}
 \resizebox{8.5cm}{!}{\includegraphics[angle=0]{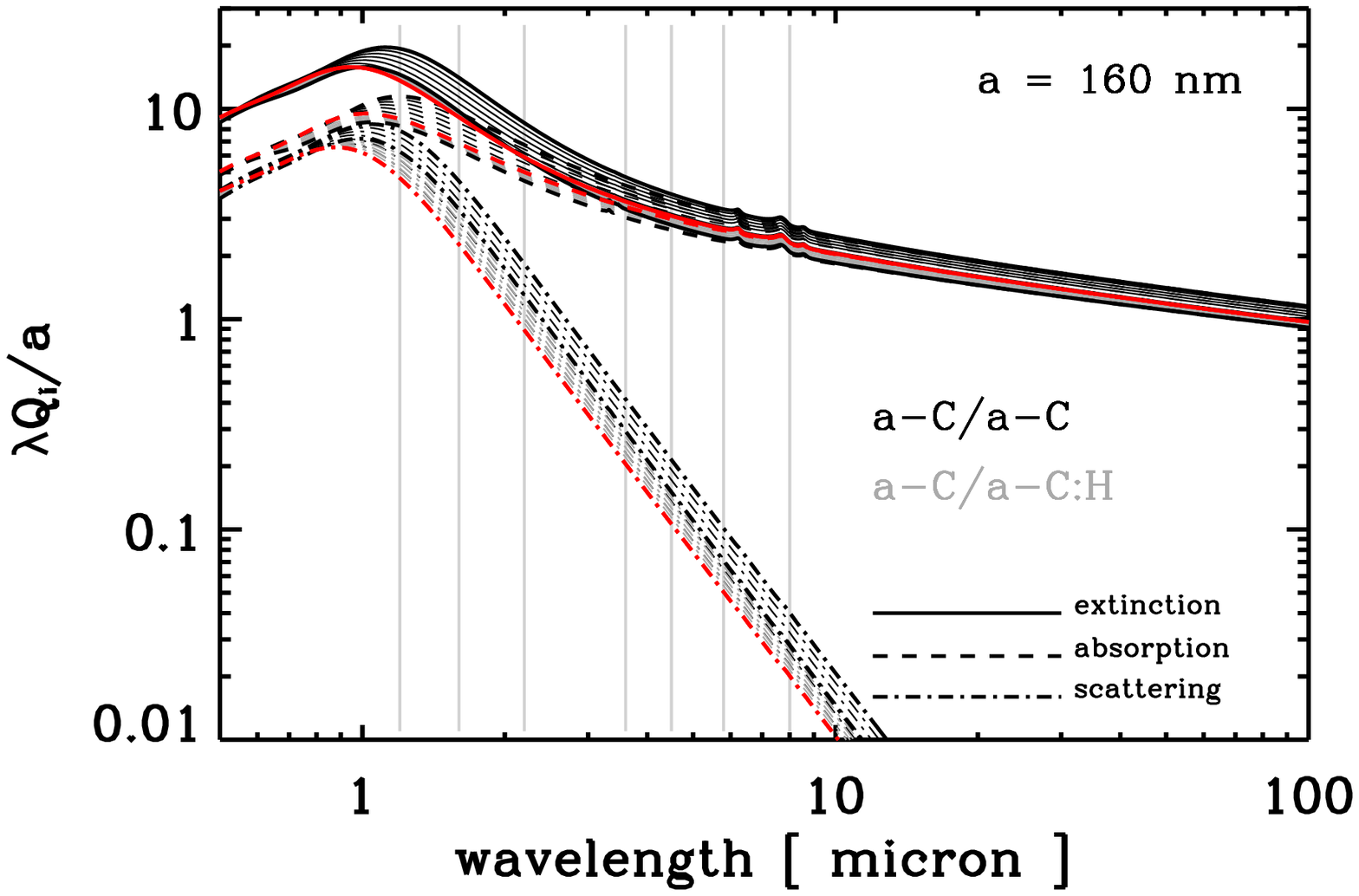}}
 \caption{$\lambda Q_{i}/a$ (where $i=$ ext, abs and sca) {\it vs.} wavelength as a function of the mantle depth, $d$, increasing away from 0\,nm (red) to 30\,nm (black) in steps of 5\,nm for each core/mantle combination. a-Sil(ol.) indicates an olivine-type amorphous silicate composition. The vertical grey lines indicate the nominal J, H, K and IRAC 3.6, 4.5, 5.8 and 8\,$\mu$m photometric band positions.}
 \label{fig_Qexscab}
\end{figure}

In Fig.~\ref{fig_Qexscab} we show the core/mantle grain extinction efficiency factors $Q_{i}$ (for $i = $ ext, sca and abs) plotted as $\lambda Q_{i}/a$, where $a$ is the grain radius, for a-C and a-C:H mantled olivine-type a-Sil$_{\rm Fe,FeS}$ cores (upper panel) and a-C cores (lower panel) as a function of the mantle depth, $d$; the red lines show the data for the bare cores. This figure clearly shows that strongly-absorbing a-C mantles on weakly-absorbing cores ({\it e.g.}, a-Sil$_{\rm Fe,FeS}$) significantly increase the absorption in the $1-8\,\mu$m region, while little affecting the scattering properties. However, adding grain mass in the form of weakly-absorbing a-C:H mantles leads to only small changes in the optical properties right across the spectrum ($\lambda = 0.5-100\,\mu$m). In fact, the addition of a-C:H mantles increases $Q_{\rm sca}$ but decreases $Q_{\rm abs}$ at these wavelengths. The net result is that $Q_{\rm ext}$ increases (decreases) for $\lambda \lesssim 3\,\mu$m ($\lambda \gtrsim 3\,\mu$m).\footnote{The data for pyroxene-type cores with Fe/FeS nano-inclusions are very similar to those for olivine-type cores (see Fig. \ref{fig_app_Qexscab}, Appendix \ref{appendix_spare_figs}).} The effects of both a-C and a-C:H mantles on strongly-absorbing grain cores ({\it e.g.}, a-C) are much less marked (see Fig.~\ref{fig_Qexscab}, lower panel). 

We find that a-C mantles lead to a slight decrease in the albedo ($Q_{\rm sca}$/$Q_{\rm ext}$), while a-C:H mantles on the contrary lead to a slight increase in the albedo (see Fig. \ref{fig_app_albedos}, Appendix \ref{appendix_spare_figs}).

\begin{figure}
 \resizebox{8.5cm}{!}{\includegraphics[angle=0]{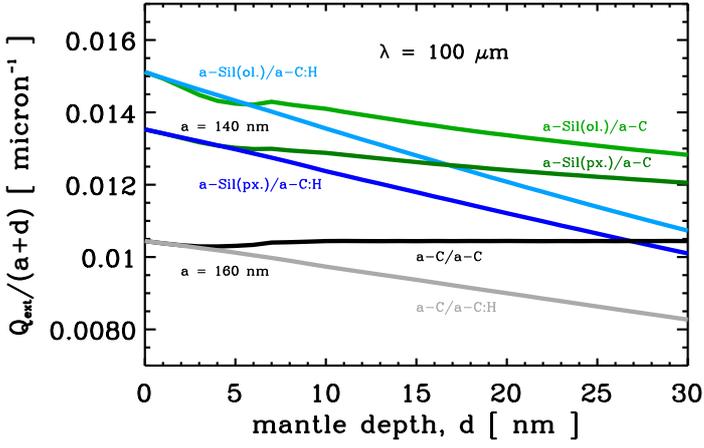}}
 \caption{$Q_{\rm ext}/(a+d)$ {\it vs.} mantle depth, $d$, at wavelength $\lambda = 100\,\mu$m, for core/mantle particles with a-C and a-C:H mantles.  a-Sil(px.) indicates a pyroxene-type amorphous silicate composition.}
 \label{fig_100mic}
\end{figure}

At long wavelengths adding mantles of the same material onto grain cores, {\it i.e.}, increasing the grain size for a homogeneous grain, leads to the expected constancy of $Q_{\rm ext}/a = Q_{\rm abs}/a$. This behaviour is seen for a-C mantles on a-C cores for mantle depths $\gtrsim10$\,nm\footnote{For mantle thicknesses $\lesssim 10$\,nm the non-horizontal behaviour is due to an increase in the effective band gap at small dimensions, see \cite{2012A&A...542A..98J}, {\it i.e.}, the mantle material band gap is $> 0.1$\,eV and the grains are core/mantle and no longer homogeneous.} (the horizontal black line in Fig.~\ref{fig_100mic}).All other core/mantle combinations do not follow the same constant $Q_{\rm ext}/a$ behaviour because the grains are not homogeneous, {\it i.e.},  the core and mantle optical properties differ; we return to this key point below. 

\begin{figure}
 \resizebox{8.5cm}{!}{\includegraphics[angle=0]{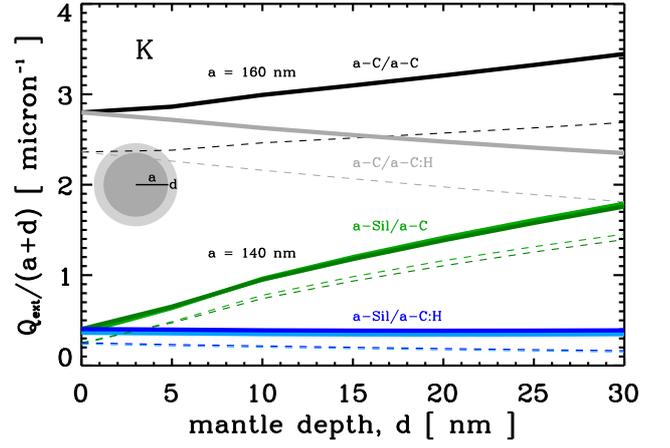}}
 \caption{Same grain data and colour-coding scheme as for Fig.~\ref{fig_100mic} but at the photometric K band wavelength (2.2\,$\mu$m). The thin dashed lines show the equivalent for the absorption efficiency factor $Q_{\rm abs}/(a+d)$. In the inset core/mantle schematic the mantle thickness has been exaggerated for clarity.} 
 \label{fig_Qad_K}
\end{figure}

\begin{figure}
 \resizebox{8.5cm}{!}{\includegraphics[angle=0]{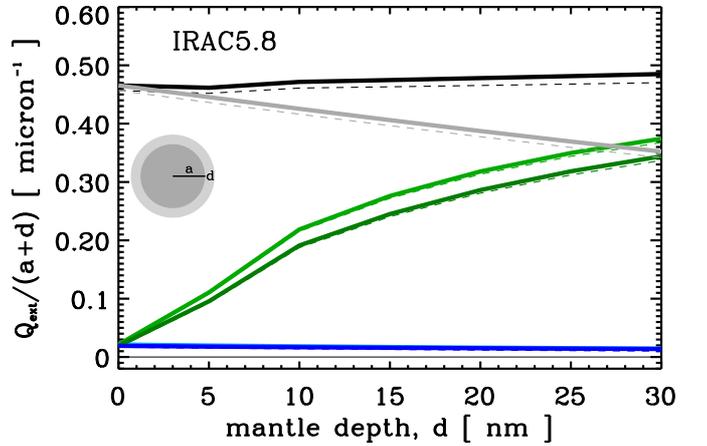}}
 \caption{Same grain data as for Figs.~\ref{fig_100mic} and \ref{fig_Qad_K} but for the IRAC 5.8\,$\mu$m band.}
 \label{fig_Qad_IRAC2}
\end{figure}

In the C-shine wavelength regime ($\lambda \simeq 1 - 8 \,\mu$m) $Q_{\rm ext}/a$ is not independent of $a$ and in Figs.~\ref{fig_Qad_K} and \ref{fig_Qad_IRAC2} we show the predicted $Q_{\rm ext}/(a+d)$ {\it vs.} mantle depth, $d$, behaviour at the photometric K and IRAC 5.8\,$\mu$m bands for olivine- and pyroxene type a-Sil$_{\rm Fe,FeS}$ and a-C grain cores with a-C and a-C:H mantles. 
The first thing to note in these figures is that the addition of absorbing a-C mantles on all grain core types leads to an increase in $Q_{\rm ext}/(a+d)$ as $d$ increases.\footnote{The `knee' in the data for a mantle thickness $\sim 7$\,nm is due to the larger effective band gap (see the above footnote).} Secondly, the addition of a-C:H mantles on a-Sil$_{\rm Fe,FeS}$ cores leads to approximately constant values of $Q_{\rm ext}/(a+d)$, while this ratio actually decreases with increasing $d$ for a-C:H mantles on a-C cores. 
Thirdly, we note that the optical properties of these materials in the K and IRAC 5.8\,$\mu$m bands are not due to pure absorption, an effect that is more pronounced in the K band data.\footnote{The data for the photometric J and H  bands and the IRAC 3.6, 4.5 and 8\,$\mu$m bands follow the same trends as for the K and IRAC 5.8\,$\mu$m bands (see Fig. \ref{fig_app_Qad_JHK}, Appendix \ref{appendix_spare_figs}).} 

\begin{figure}
 \resizebox{8.5cm}{!}{\includegraphics[angle=0]{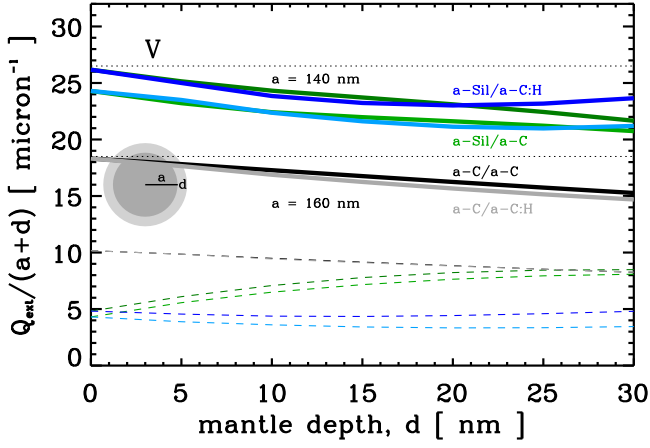}}
 \caption{Same grain data and colour-coding scheme as for Fig.~\ref{fig_100mic} but at the  V band wavelength (0.54\,$\mu$m). The thin dashed lines show the equivalent for the absorption efficiency factor $Q_{\rm abs}/(a+d)$. The horizontal dotted lines are for visual reference only.}
 \label{fig_Qad_V}
\end{figure}

In Fig.~\ref{fig_Qad_V} we show the effects of a-C and a-C:H mantles on grains in the visible (V band). Here we see that, for all mantle and core material combinations, the extinction cross-sections decrease slightly with increasing particle radius, {\it i.e.}, with increasing dust mass. 

These data therefore reveal a rather counter-intuitive scenario where adding an a-C:H mantle onto dust ({\it i.e.}, adding dust mass) actually leads to no change or even a decrease in the dust extinction cross-sections in the V band and in the $1-8\,\mu$m wavelength region, which are accompanied by an increase in the dust albedo. This strange behaviour is a direct consequence of the optical properties of wide band gap, amorphous, semiconducting a-C:H solids.

\begin{figure}
 \resizebox{\hsize}{!}{\includegraphics{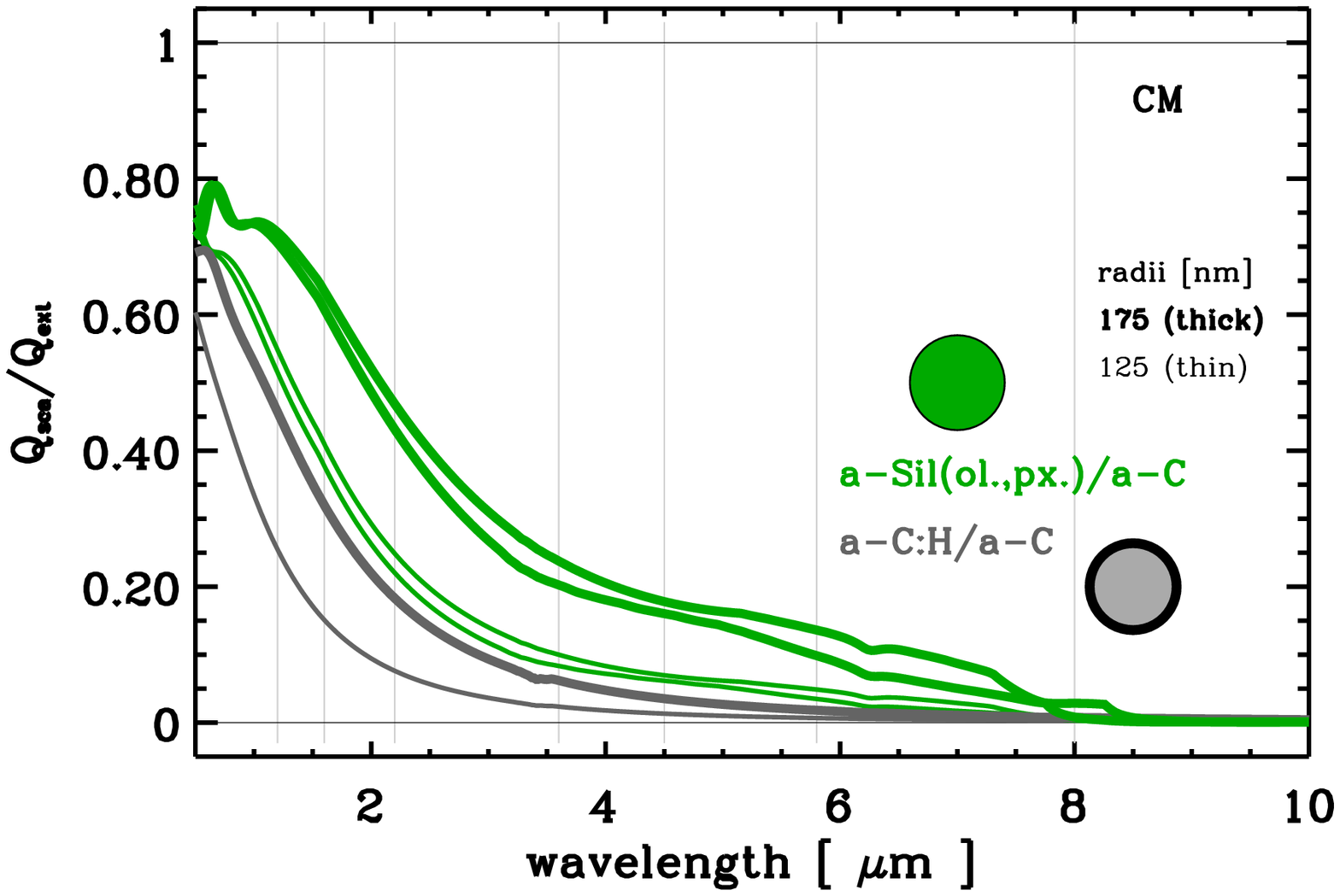}}
 \resizebox{\hsize}{!}{\includegraphics{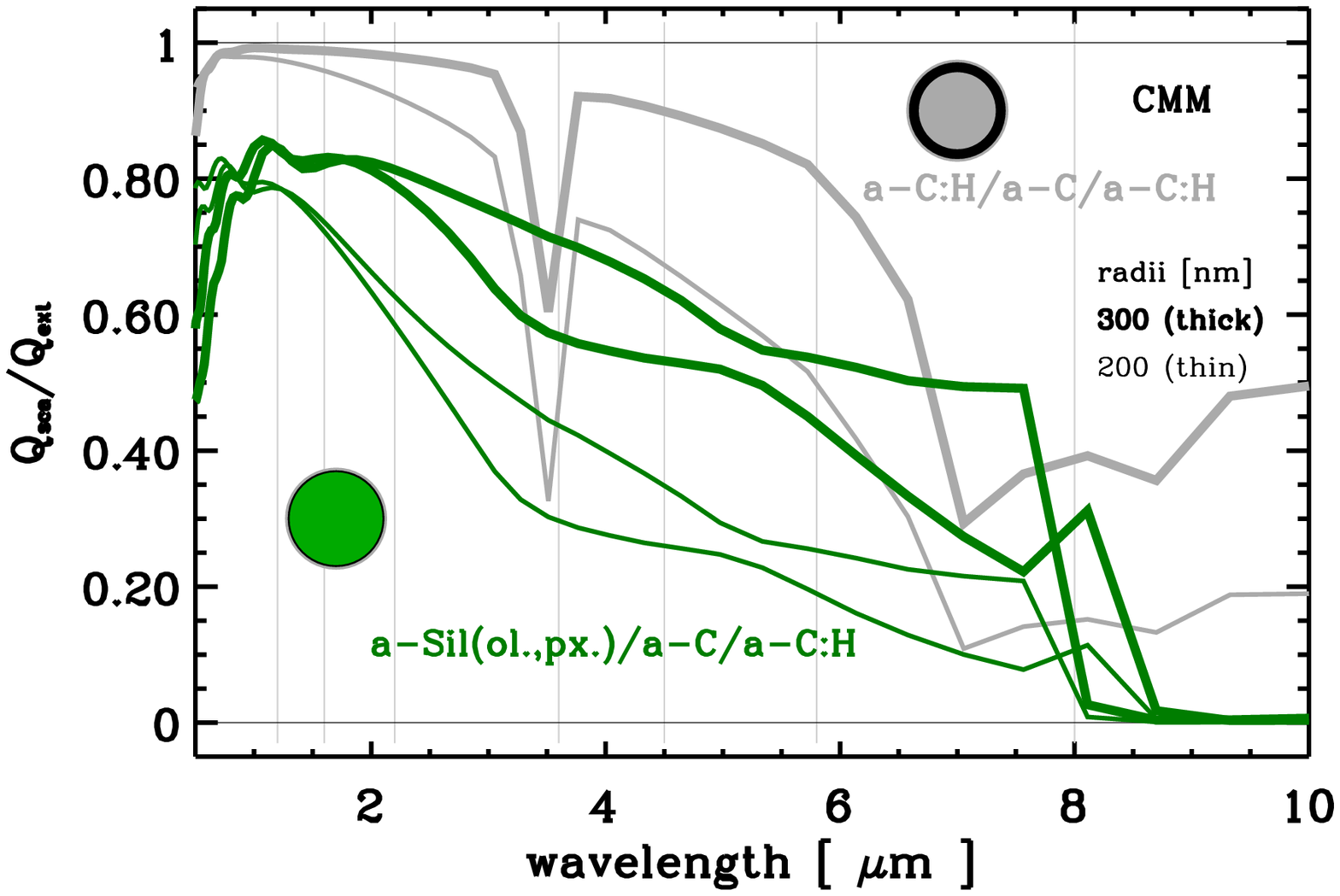}}
 \resizebox{\hsize}{!}{\includegraphics{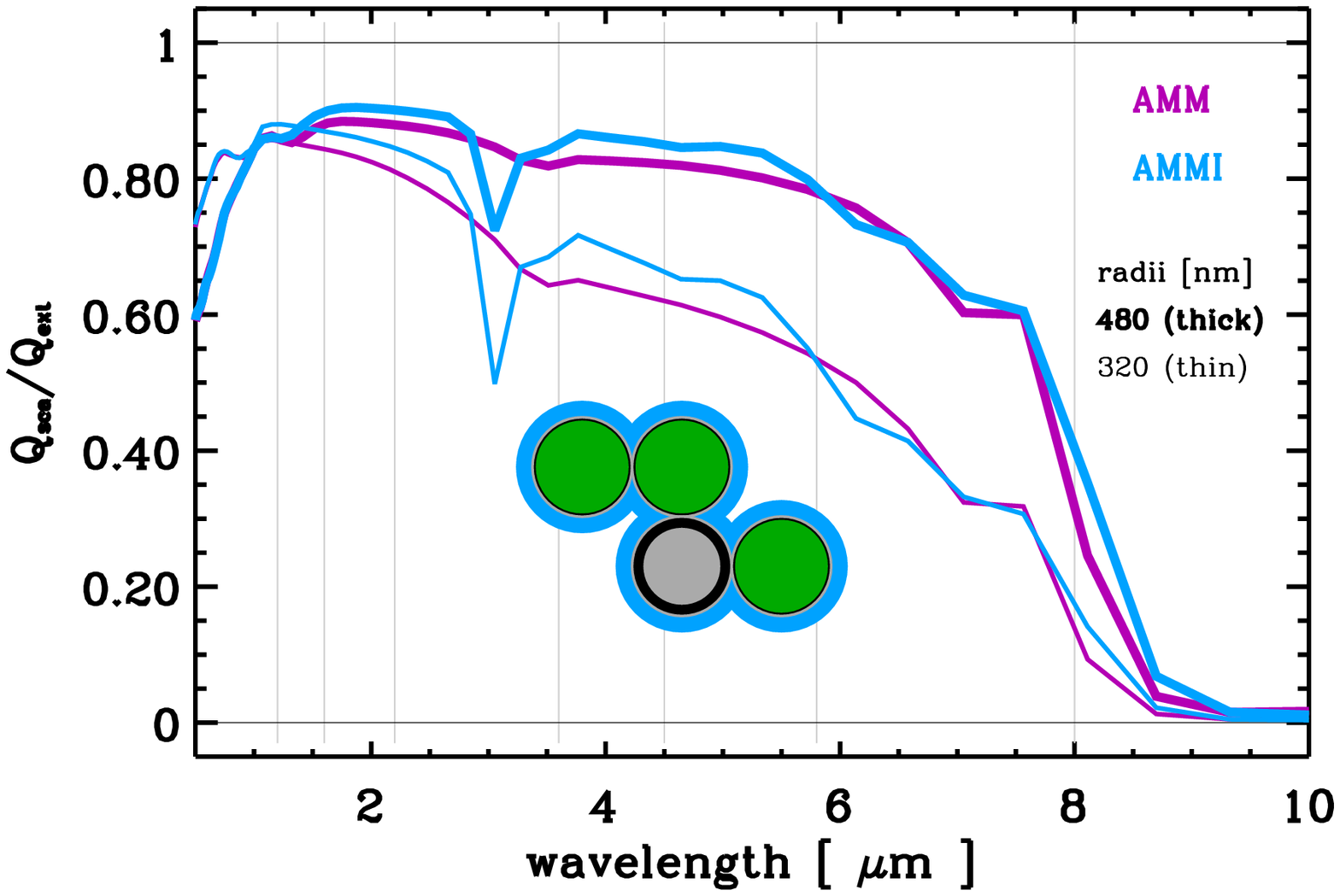}}
 \caption{The \cite{2013A&A...558A..62J} and \cite{2014A&A...565L...9K,2015A&A...0000...000} dust model albedos plotted as $Q_{\rm sca}/Q_{\rm ext}$, for volume-equivalent grain radii near the mass distribution peaks for CM (top), CMM (middle), AMM and AMMI grains (bottom). In the grain structure schematics the core radii and the mantle thicknesses are approximately to scale. Note that the wavelength resolution is lower than in Fig.~\ref{fig_QsQe}.}
 \label{fig_QsQe_model}
\end{figure}

\subsection{Towards a more realistic dust evolution model: THEMIS} 
\label{sect_anderson_loc_cmmamm}

Here we investigate mantle effects within the framework of the multi-component and aggregate grains (consisting of a-Sil$_{\rm Fe,FeS}$, a-C and a-C:H materials), with more complex, physically-realistic grain structures. In this case the optical properties are calculated using the discrete-dipole approximation (DDA) method \citep{2000ascl.soft08001D} as per  \cite{2011A&A...528A..96K,2012A&A...548A..61K,2014A&A...565L...9K,2015A&A...0000...000}. As per \cite{2015A&A...0000...000} we consider the following grain and aggregate structures:
\begin{itemize}
  \item CM: core/mantle a-Sil and  a-C:H cores with a-C mantles
  \item CMM: core/mantle/mantle a-Sil and  a-C:H cores with \\ inner a-C and outer a-C:H mantles
  \item AMM: aggregates of CMM grains, and 
  \item AMMI: aggregates of CMM grains with water ice mantles.
\end{itemize} 

\begin{figure}
 \resizebox{8.5cm}{!}{\includegraphics[angle=0]{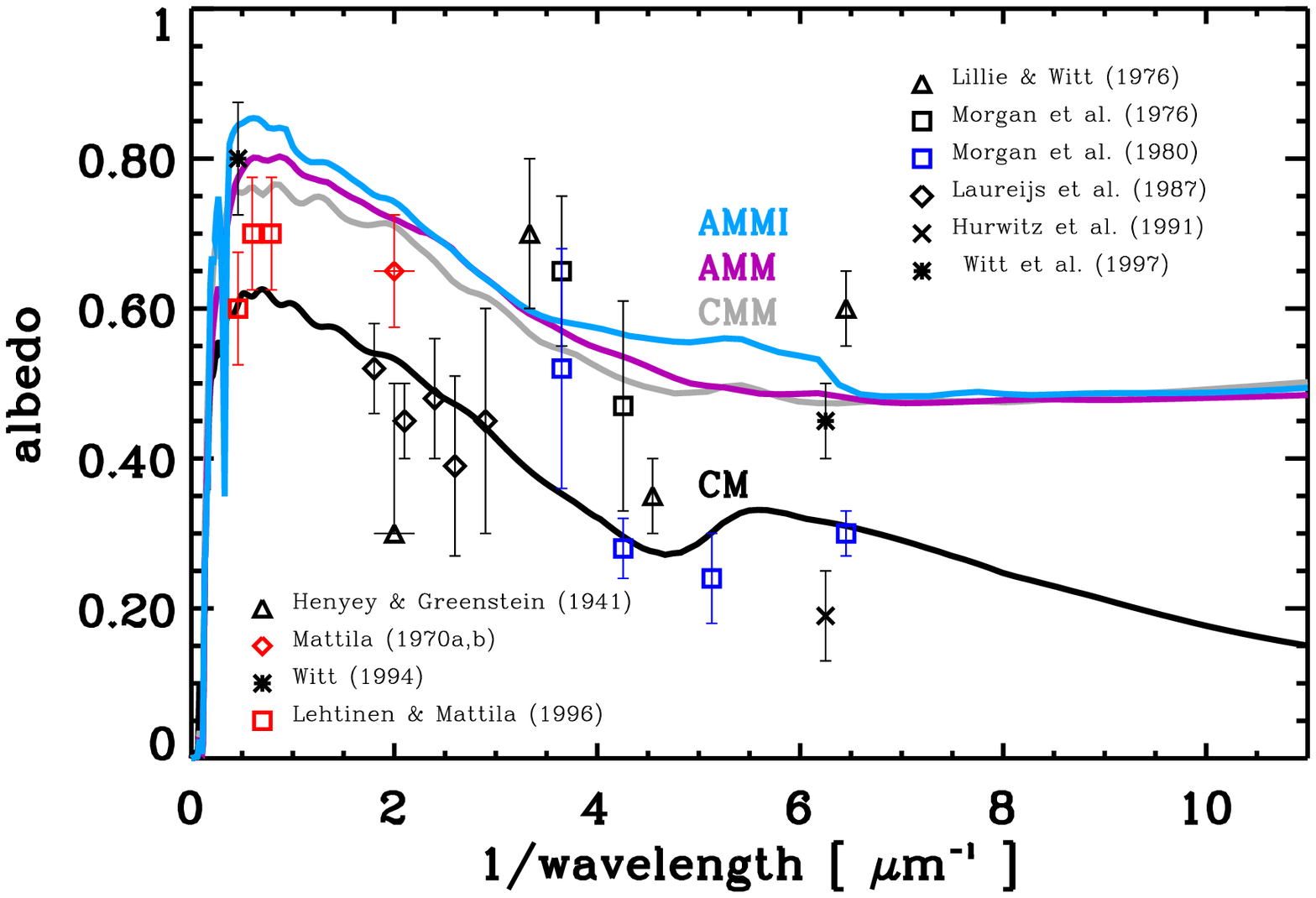}}
 \caption{The THEMIS albedos for full size distributions of CM, CMM, AMM and AMMI grains \citep{2015A&A...0000...000} compared with the available observations.  
The data points are for regions towards dark clouds \citep[red;][]{1970A&A.....8..273M,1970A&A.....9...53M,1996A&A...309..570L}, the $\lambda$~Ori dark nebular shell \citep[blue;][]{1980MNRAS.190..825M} and for the diffuse galactic light \citep[DGL, black;][]{1941ApJ....93...70H,1976ApJ...208...64L,1976MNRAS.177..531M,1987A&A...184..269L,1991ApJ...372..167H,1994ApJ...427..227W,1997ApJ...481..809W}.} 
 \label{fig_albedo}
\end{figure}

In general, the complexity of the dust composition and structure increases with increasing density along the sequence CM $\rightarrow$ CMM $\rightarrow$ AMM $\rightarrow$ AMMI. The THEMIS albedos are shown in Fig. \ref{fig_QsQe_model}, plotted as $Q_{\rm sca}/Q_{\rm ext}$, for grain radii near the mass distribution peaks for the CM (upper panel), CMM (middle panel), and AMM and AMMI (lower panel) grains. We note that the addition of a-C:H mantles in the CM~$\rightarrow$~CMM transition leads to enhanced scattering, especially in the case of the aC:H/a-C/a-C:H CMM grains (grey lines in the middle panel). These effects become compounded when the CMM grains are coagulated into aggregates (lower panel). We find that the condition $Q_{\rm sca} >  Q_{\rm abs}$ for CM grains occurs only in the $\lambda \lesssim 1\,\mu$m region, whereas for CMM (AMM/AMMI) grains this occurs for $\lambda < 5\,\mu$m ($\lambda < 8\,\mu$m) as shown in more detail in Fig.~\ref{fig_Qs_model} in Appendix~\ref{appendix_spare_figs}.

In Fig.~\ref{fig_albedo} we show the THEMIS albedos \cite[for the full size distributions of CM, CMM, AMM and AMMI grains,][]{2015A&A...0000...000} compared with the available observational data for dark clouds (red and blue data points) and the diffuse galactic light (DGL, black data points). 
From these data we note that the CMM, AMM and AMMI models are consistent with the NIR dust albedos \citep[J, H and K,][]{1996A&A...309..570L} and the visible UV albedos \citep[$1/\lambda < 4\,\mu$m$^{-1}$,][]{1970A&A.....8..273M,1970A&A.....9...53M,1980MNRAS.190..825M} for dark clouds (red and blue data points), while the standard diffuse ISM model matches the DGL data more closely. 
At shorter wavelengths ($\lambda < 250$\,nm) there appears to be an inherently greater spread in the observed albedos. The THEMIS albedos for CM, CMM, AMM and AMMI grains cover approximately this same range. However, the number of albedo determinations is limited because the measurements are fraught with difficulty and the uncertainties are therefore generally rather large. We note that the most recent of the albedo determinations shown in Fig.~\ref{fig_albedo} are almost 20 years old. 

We have also investigated the effects of coagulating CM grains into aggregates and find that their optical properties (absorption, scattering, albedo {\it etc.}) lie closer to those of the CM grains than either the CMM or AMM/AMMI grains. 
Hence, the coagulation of CM grains into aggregates results in optical properties that are not compatible with the observed evolution of the dust properties in the diffuse to dense cloud transition \citep{2015A&A...0000...000}.

\section{Astrophysical implications and consequences} 
\label{sect_implications}

The dust evolution processes that we have explored here and elsewhere \citep[{\it e.g.},][{\it etc.}]{2013A&A...558A..62J,Faraday_Disc_paper_2014,2014A&A...565L...9K,2015A&A...Ysard_etal,2015A&A...0000...000}, would appear to have some important consequences and wider implications for dust and dust-related chemistry in the diffuse to dense cloud transition. 

As discussed by \cite{2013A&A...558A..62J,Faraday_Disc_paper_2014}, carbon accretion and the formation of a-C(:H) mantles appears to be consistent with the observed carbon depletion and allied variations in the extinction curve \citep{2012ApJ...760...36P,2013A&A...558A..62J}. It now appears that carbon could be more abundant in the ISM than previously assumed \citep[{\it e.g.}, $\simeq 355\pm 64$\,ppm,][]{2012ApJ...760...36P}, which if we assume that $\simeq 200$\,ppm are in dust in the diffuse ISM \citep{2013A&A...558A..62J,2015A&A...Ysard_etal} would leave $\simeq 155 \pm 64$\,ppm in the gas in the form of CO (and other molecules) and available for carbonaceous mantle formation in the transition from the diffuse ISM to denser regions \citep{2015A&A...0000...000}. This is consistent with dust evolution in the outer regions of dense clouds, through a-C:H mantle formation and small grain coagulation onto larger grains, resulting in extinction that is characterised by a weak UV bump and a steep rise in the FUV extinction, \citep[{\it c.f.}, the HD 207198 line of sight;][]{2012ApJ...760...36P}.

Grain surface-driven chemistry and ice accretion have a long and intertwined history. Our work indicates that the surface chemistry in denser interstellar regions (in the absence of icy mantles) probably occurs on a-C:H surfaces that could perhaps play an active role in the chemistry. Thus, the nature of the surface chemistry and molecule formation in translucent regions of the ISM may need to be carefully re-evaluated. 

The addition of dust mass via the formation of aliphatic-rich mantles leads to no change in, or even a decrease in, the dust extinction cross-section for large ($a = 100 - 300$\,nm) grains at wavelengths $\simeq 0.5 - 8\,\mu$m. 
The J, H and K photometric band and Spitzer-IRAC observations are often used to characterise dust at these near-IR wavelengths. In particular the K band is generally the extinction reference normalisation of choice for low extinction and molecular clouds regions. Our results suggest that extreme caution should be exercised in the derivation of any dust sizes and masses from the shape and absolute values of J-, H- or K-normalised near-IR extinction data. Additionally, in translucent clouds the value of $A_{\rm V}$ derived by extrapolation from J, H and K photometric data is likely to be unreliable. 

\section{Conclusions}
\label{sect_conclusions}

We propose that, within the framework of our THEMIS dust modelling approach, the C-shine observations (cloud- and core-shine) are likely due to the combined effects of aliphatic-rich ({\it i.e.}, wide band gap a-C:H) hydrocarbon mantle formation and dust coagulation in the diffuse to dense ISM transition. These dust mass- and size-increasing processes lead to a decrease in dust extinction (absorption) cross-section and therefore a net  increase in dust albedo because the dust scattering properties are not significantly affected. The counter-intuitive combination of decreased extinction (absorption) with increased dust size and mass is a peculiarity of the optical properties of a-C:H materials. This effect is due to the very low absorptive index, $k$, but relatively high real part, $n$, of the complex index of refraction, $m(n,k)$, of the a-C:H mantle materials in the visible-IR region. This results in weak IR absorption but normal scattering by the a-C:H-mantled aggregates.  

Our results show that B and V band photometric observations of the dust properties in translucent clouds qualitatively track the dust behaviour at FIR wavelengths. However, we find that the same dust will present significantly different trends in the J, H and K photometric bands. Thus, while the J, H and K data are self-consistent, they can show opposite or no trends with the visible extinction and dust emission properties. 
Thus, the de-coupling of the dust optical properties at the B and V band and FIR wavelengths from those at the J, H and K band wavelengths, due to aliphatic mantle formation, implies that any derivation of $A_{\rm V}$, dust sizes and masses from J, H and K photometric band data is likely to be problematic and will need to be carefully re-considered in much greater detail than has been possible here. 

In the companion paper (paper II) we couple our THEMIS dust evolution modelling to radiative transfer calculations and compare the results with the observational data.


\begin{acknowledgements} 
We thank the referee for critical remarks that helped to improve the manuscript. 
This research was, in part, made possible through the EU FP7 funded project DustPedia (Grant No. 606847).  
\end{acknowledgements}


\bibliographystyle{../AandA_Macros/bibtex/aa} 
\bibliography{biblio_HAC} 

\Online


\appendix

\clearpage
\section{Additional supporting figures}
\label{appendix_spare_figs}

Here we provide the additional figures that are refereed to in the main body of the text. 
In particular, 
Fig.~\ref{fig_Qs_model} shows that for the CM grains scattering exceeds absorption in the $\lambda < 1\,\mu$m region. However, for CMM grains scattering exceeds absorption out to longer wavelengths, {\it i.e.}, $\lambda < 5\,\mu$m, and out to even longer wavelengths for the AMM/AMMI particles ($\lambda < 8\,\mu$m).

\begin{figure}
 \resizebox{\hsize}{!}{\includegraphics[angle=0]{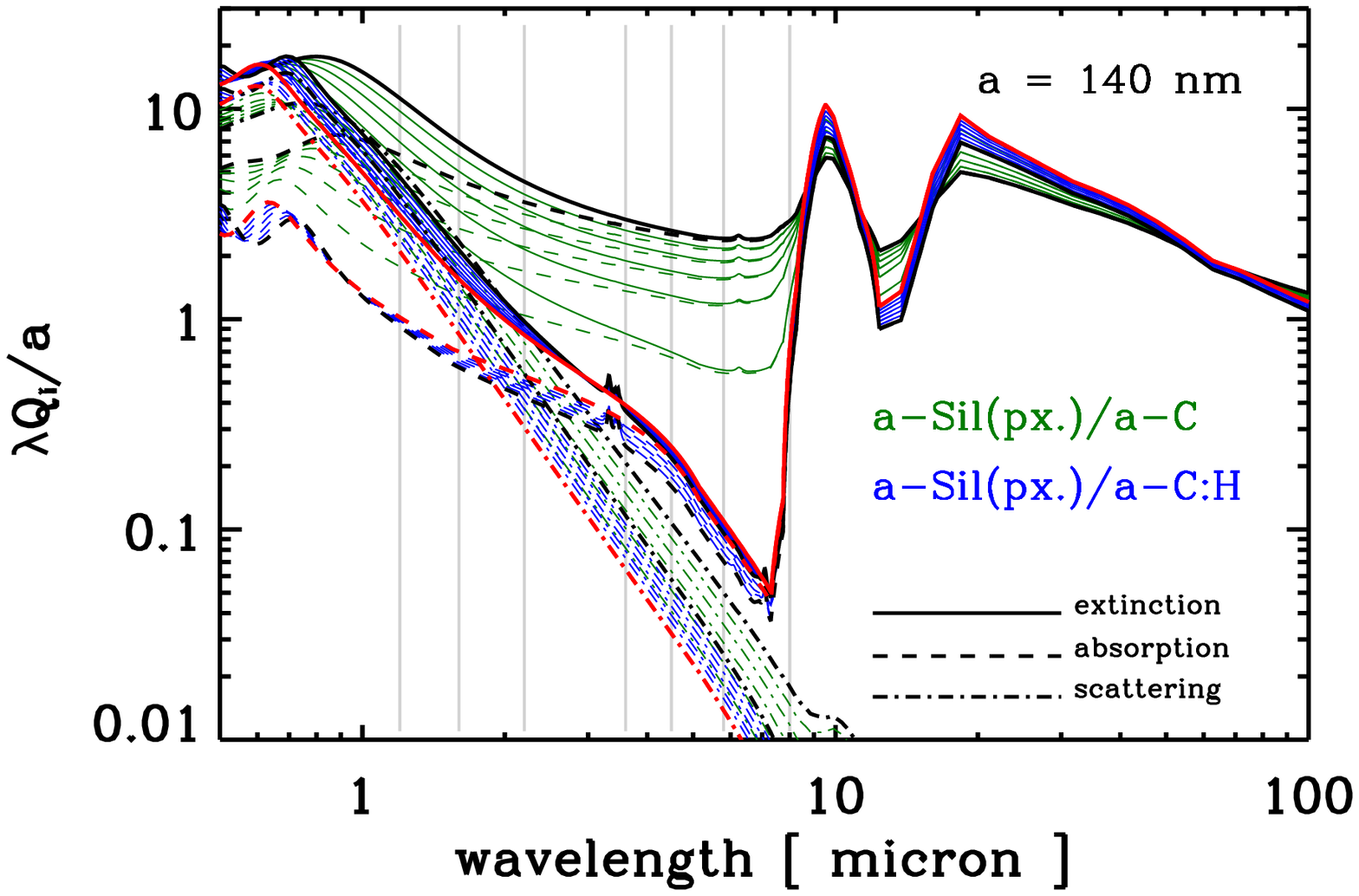}}
 \caption{The same as Fig.~\ref{fig_Qexscab} but for amorphous pyroxene-type silicate cores. $\lambda Q_{i}/a$ for (where $i=$ ext, abs and sca) {\it vs.} wavelength as a function of the mantle depth, $d$, increasing away from 0\,nm (red) to 30\,nm (black) in steps of 5\,nm, for pyroxene-type a-Sil(px.) cores mantled with a-C (dark green) and a-C:H (blue).}
 \label{fig_app_Qexscab}
\end{figure}

\begin{figure}
 \resizebox{\hsize}{!}{\includegraphics[angle=0]{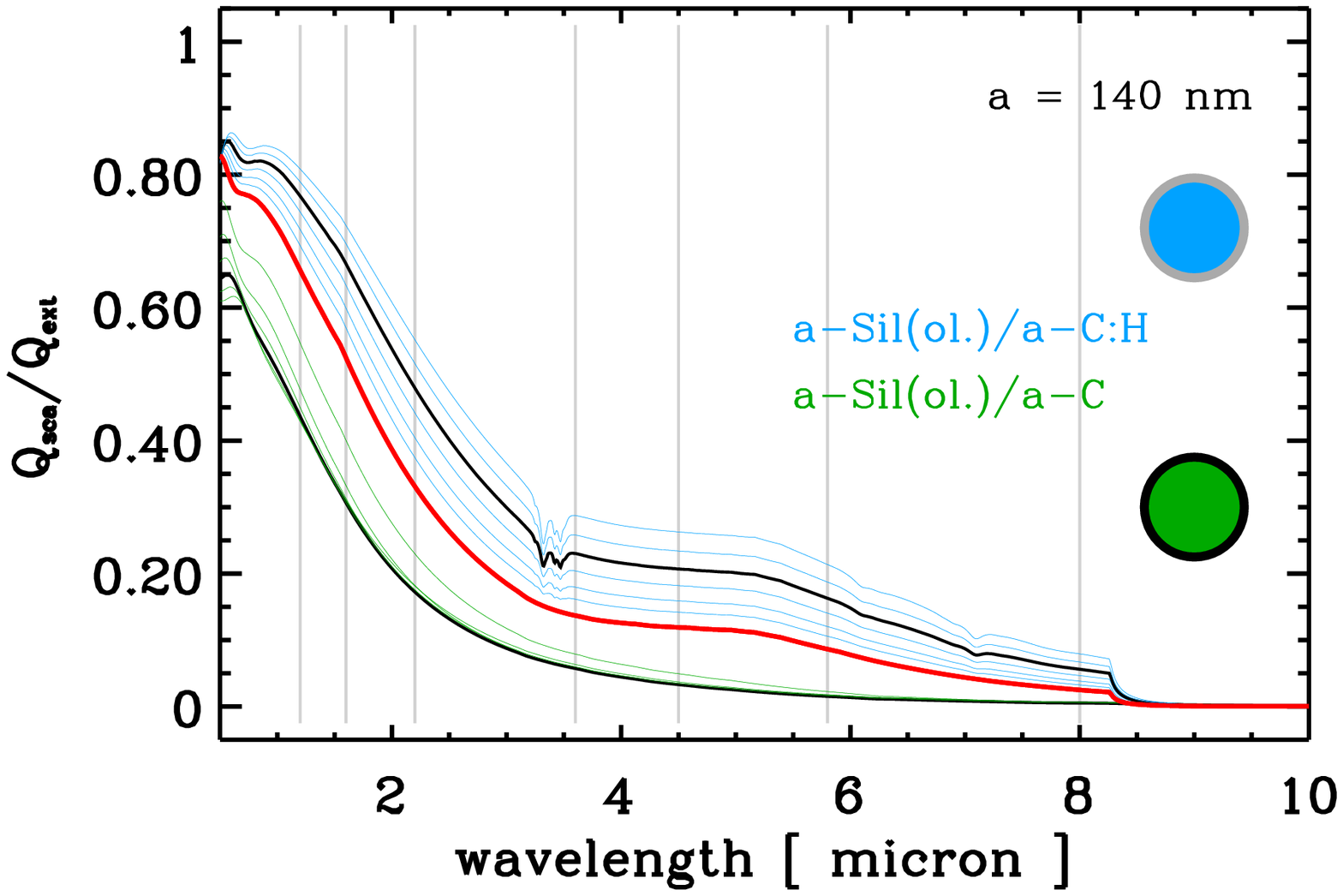}}
 \resizebox{\hsize}{!}{\includegraphics[angle=0]{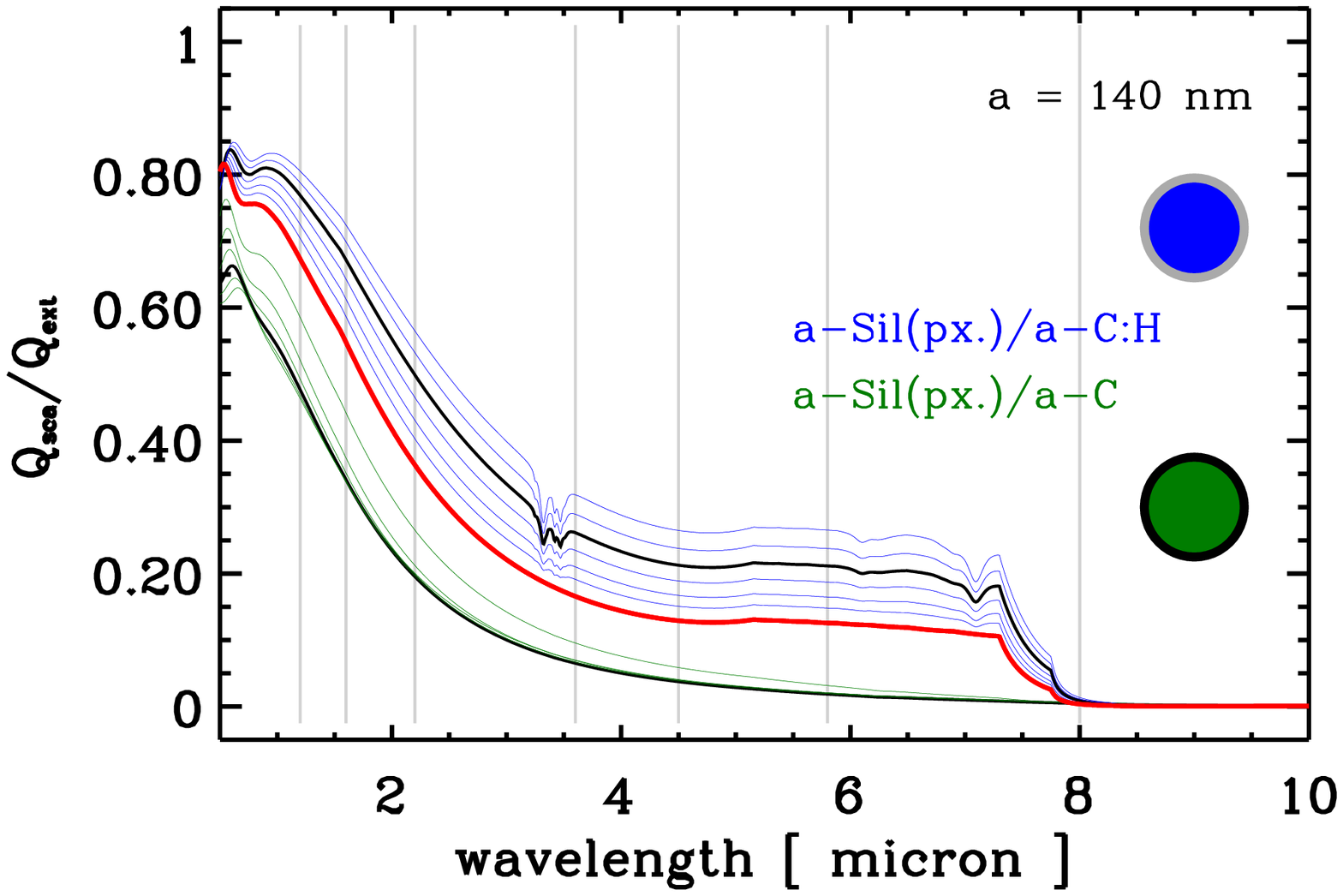}}
 \resizebox{\hsize}{!}{\includegraphics[angle=0]{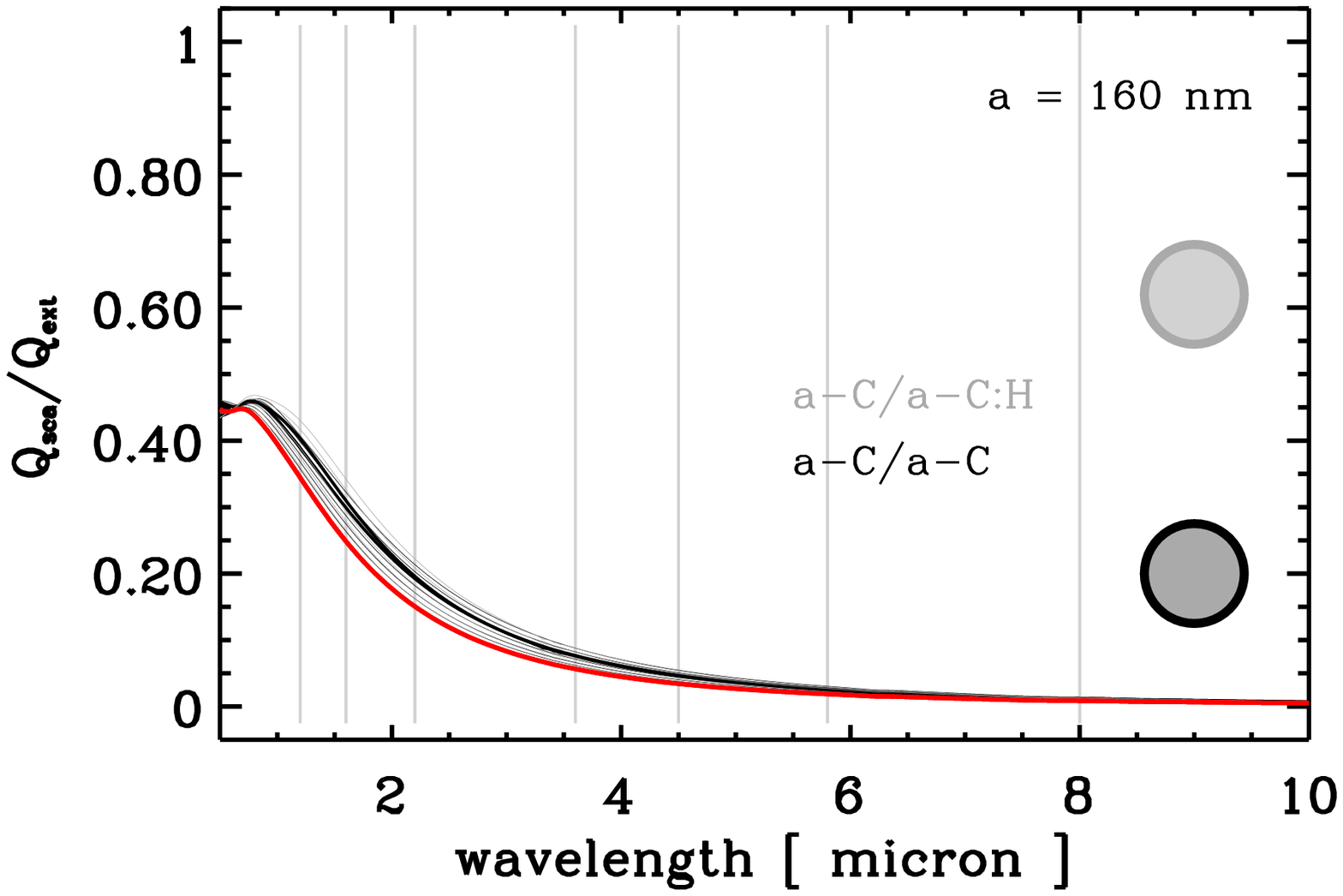}}
 \caption{$Q_{\rm sca}/Q_{\rm ext}$ {\it vs.} wavelength for each core/mantle combination as a function of the mantle depth, $d$, increasing away from 0\,nm (red) to 30\,nm (black) in steps of 5\,nm, for each core/mantle combination. The grain structure schematics show the thickest mantle case (30\,nm)  approximately to scale.}
 \label{fig_app_albedos}
\end{figure}

\begin{figure}
 \resizebox{\hsize}{!}{\includegraphics[angle=0]{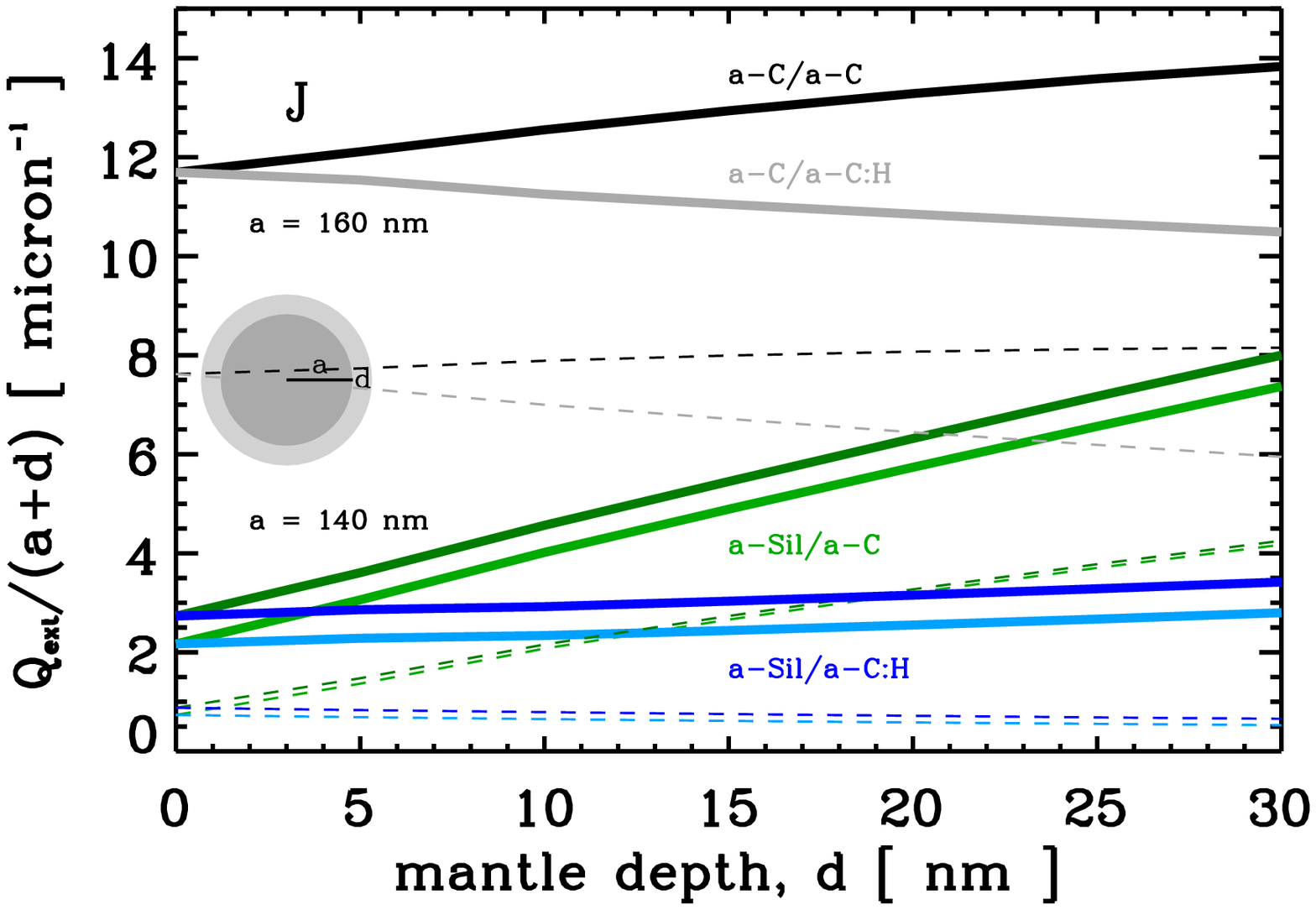}}
 \resizebox{\hsize}{!}{\includegraphics[angle=0]{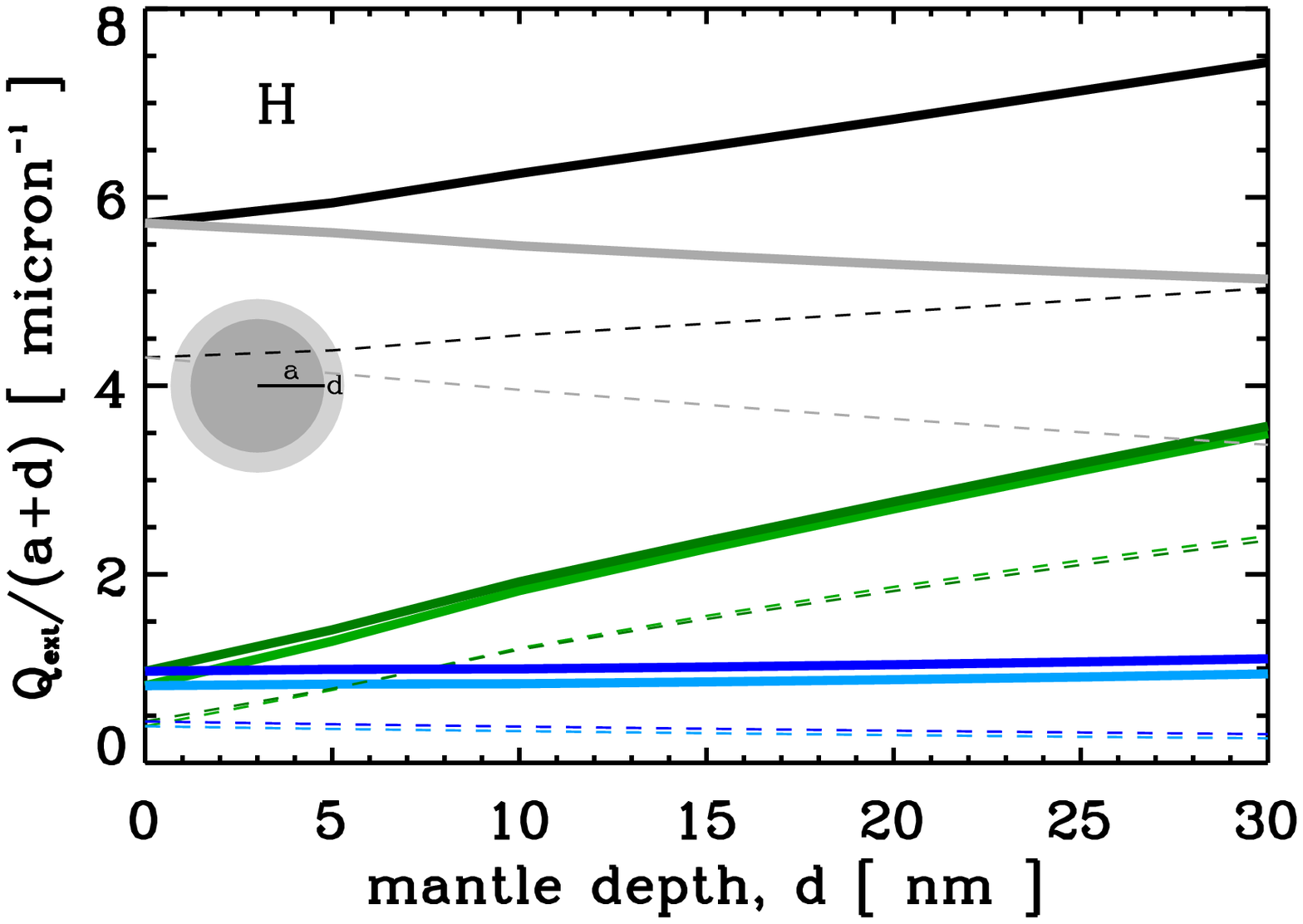}}
 \caption{$Q_{\rm ext}/(a+d)$ {\it vs.} mantle depth, $d$, for grains with a-C and a-C:H mantles at the J and H photometric band wavelengths.}
 \label{fig_app_Qad_JHK}
\end{figure}

\begin{figure}
 \resizebox{\hsize}{!}{\includegraphics[angle=0]{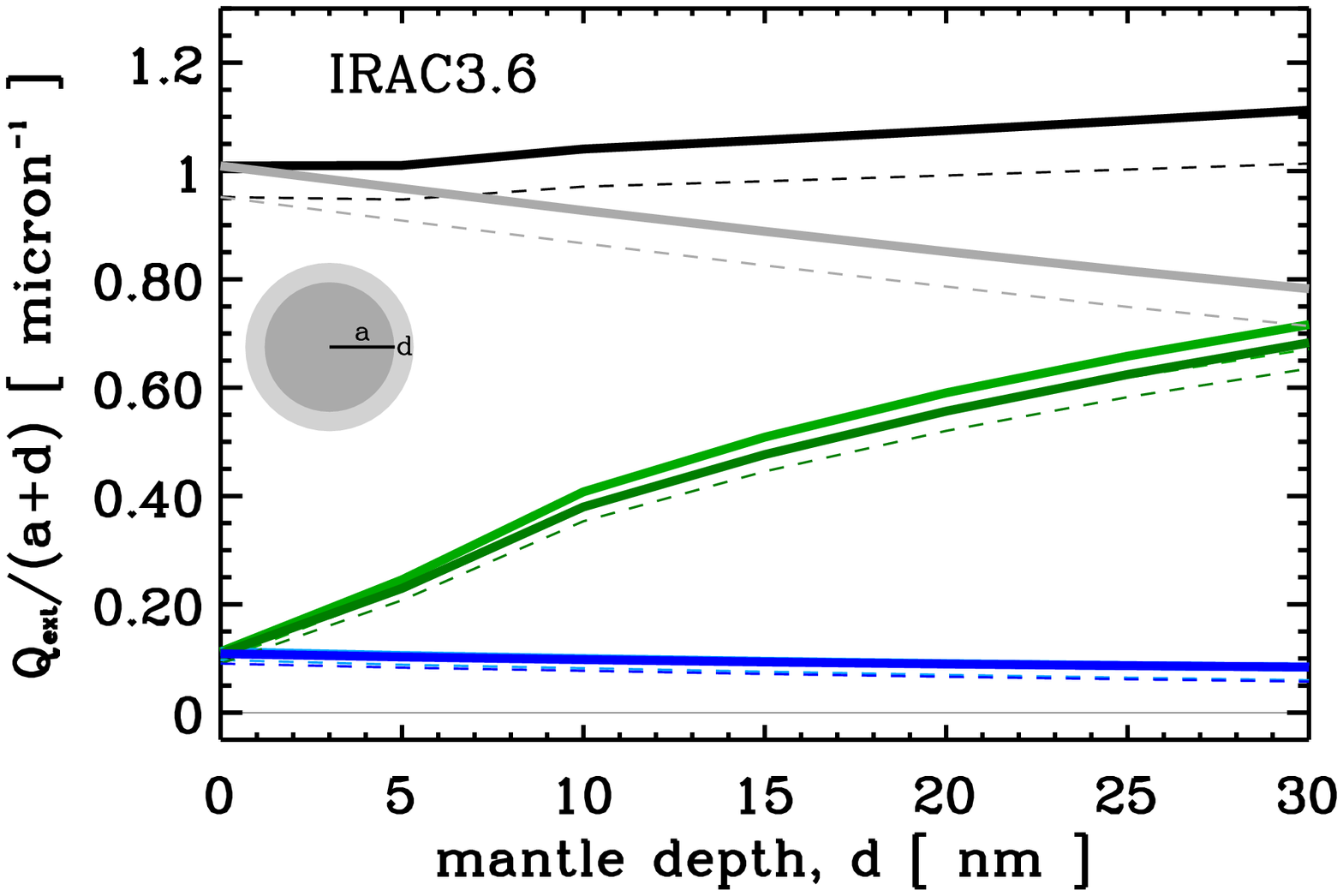}}
 \resizebox{\hsize}{!}{\includegraphics[angle=0]{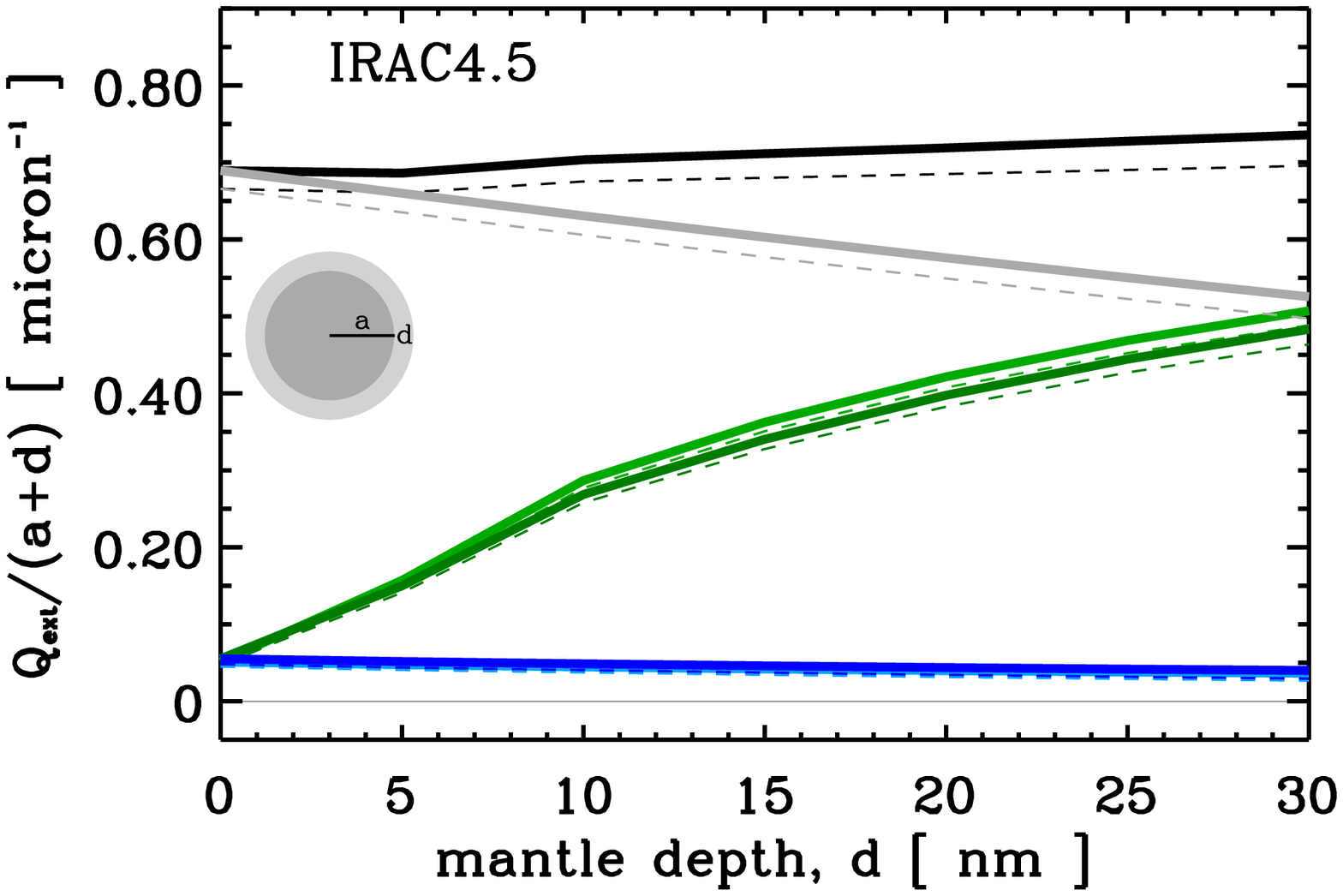}}  
 \resizebox{\hsize}{!}{\includegraphics[angle=0]{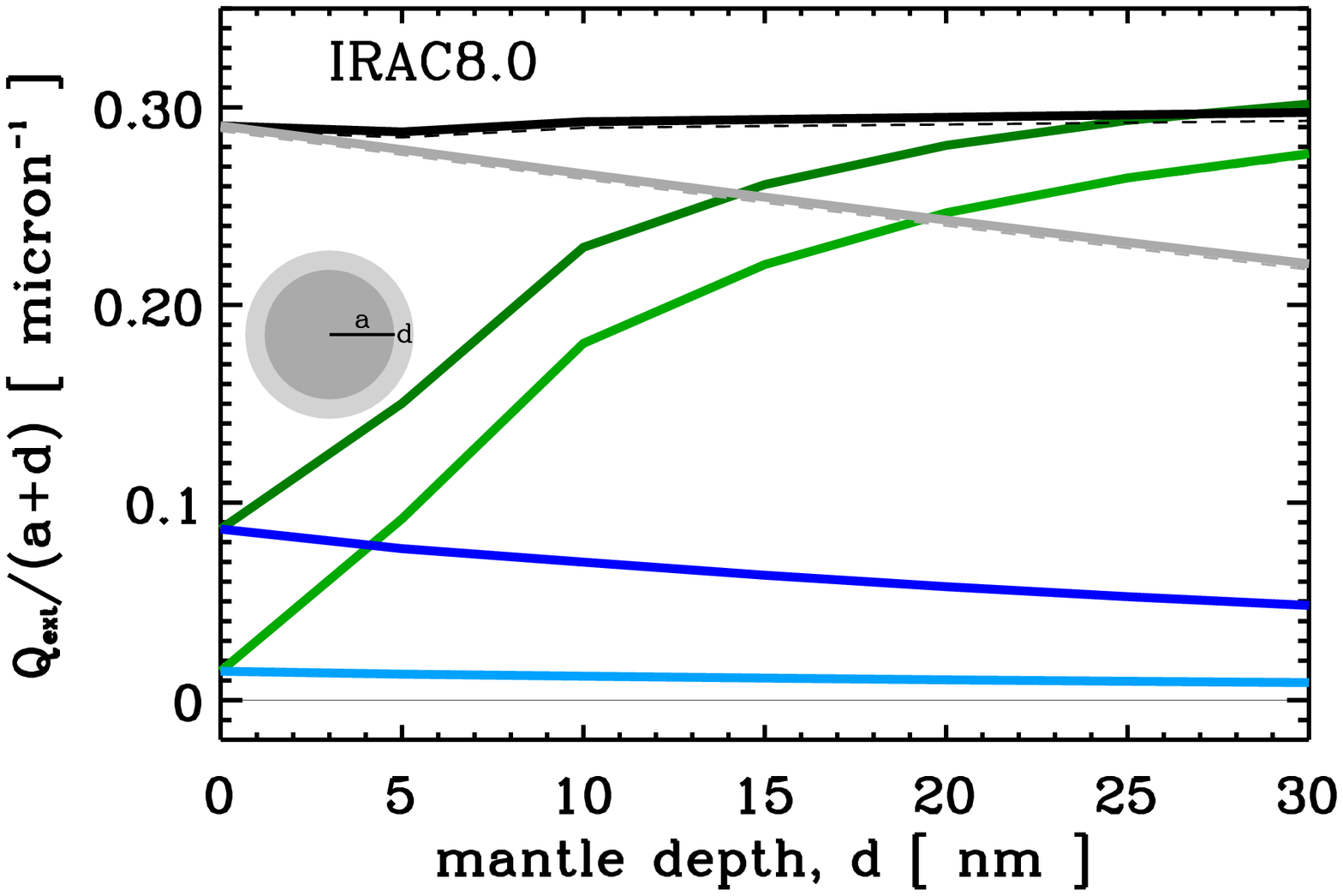}}
 \caption{$Q_{\rm ext}/(a+d)$ {\it vs.} mantle depth, $d$, for grains with a-C and a-C:H mantles at the IRAC 3.6,  4.5 and 8\,$\mu$m photometric band wavelengths.}
 \label{fig_app_Qad_IRAC1}
\end{figure}

\begin{figure}
 \resizebox{\hsize}{!}{\includegraphics{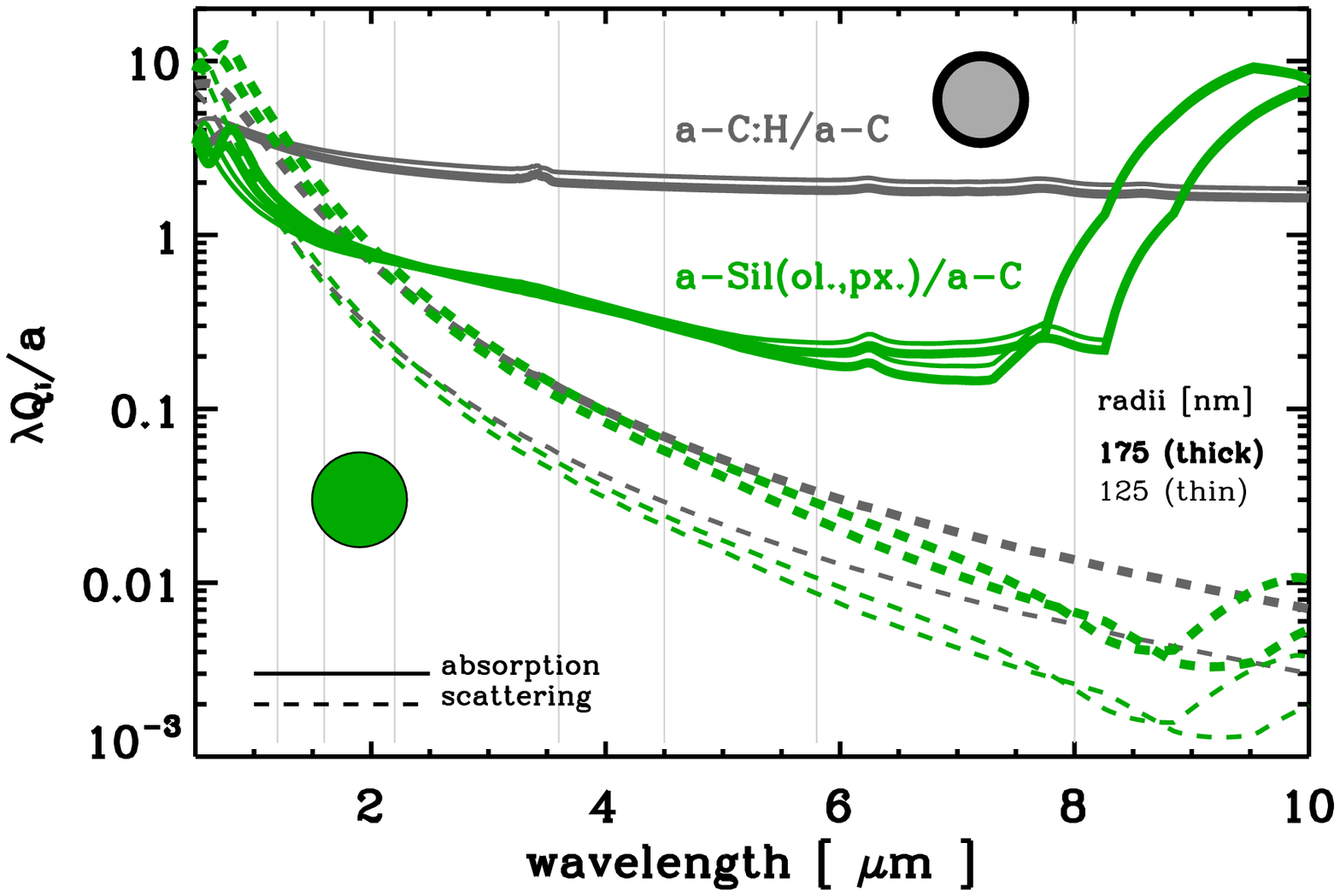}}
 \resizebox{\hsize}{!}{\includegraphics{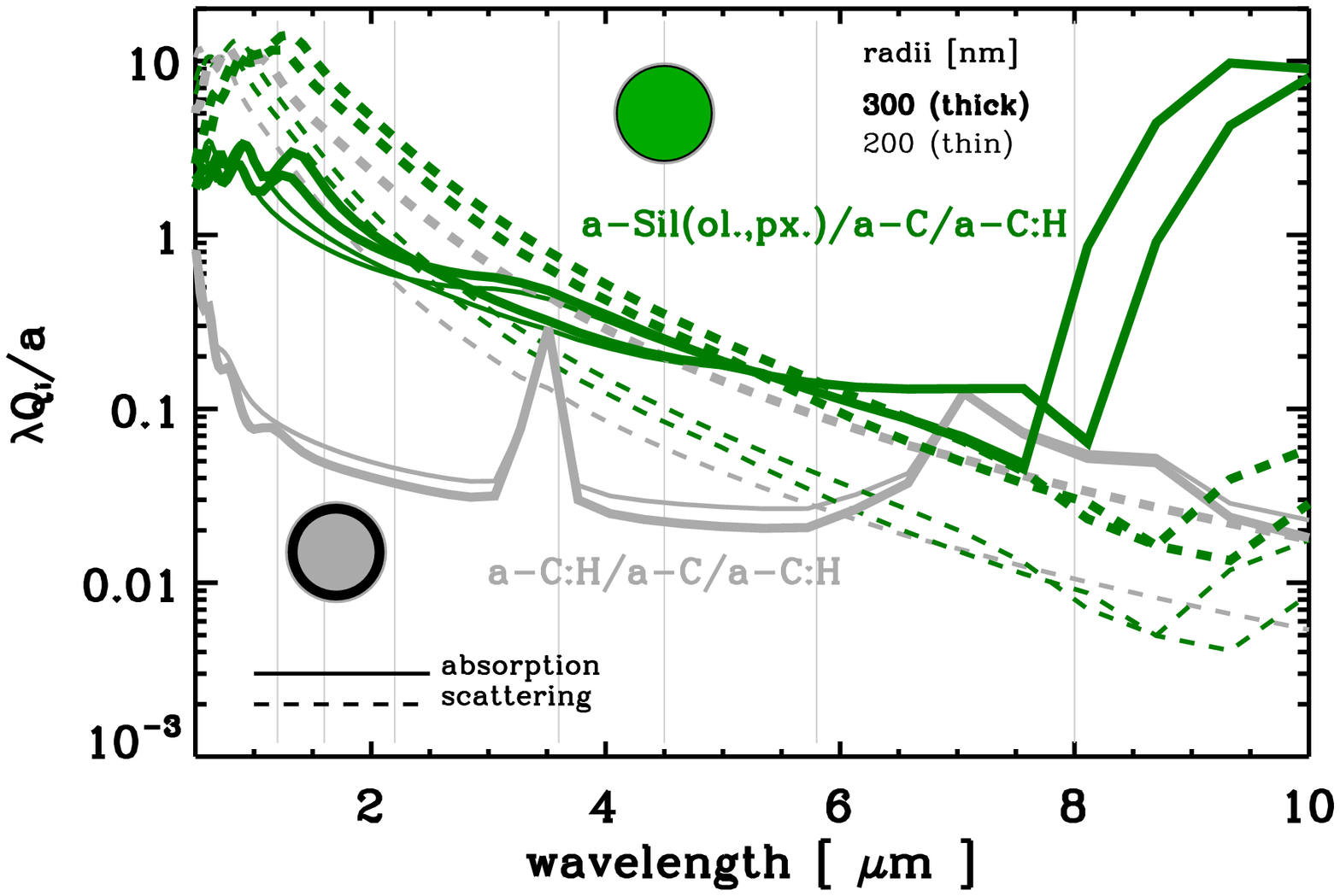}}
 \resizebox{\hsize}{!}{\includegraphics{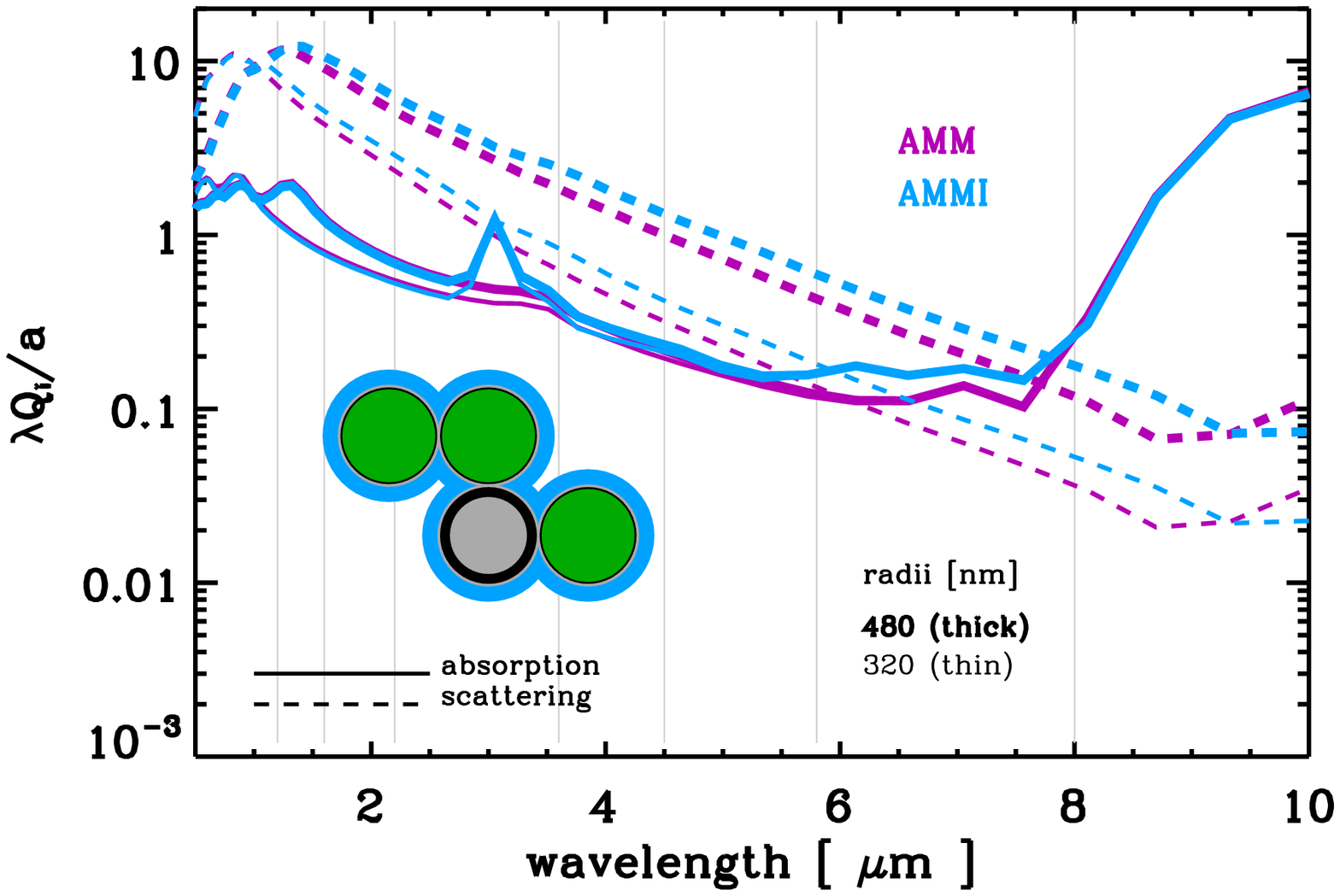}}
 \caption{The \cite{2013A&A...558A..62J} and \cite{2014A&A...565L...9K,2015A&A...0000...000} model data, for grain radii near the mass distribution peak, plotted as $\lambda Q_{i}/a$, for grain radii near the mass distribution peaks for CM (top), CMM (middle), AMM and AMMI grains (bottom). In the grain structure schematics the core radii and the mantle thicknesses are approximately to scale. The vertical grey lines indicate the J, H, K and IRAC 3.6, 4.5, 5.8 and 8\,$\mu$m photometric band positions. }
 \label{fig_Qs_model}
\end{figure}

\listofobjects

\end{document}